\documentclass[10pt,twocolumn,letterpaper]{article}

\usepackage{cvpr}              % To produce the CAMERA-READY version
\usepackage{bm}
\usepackage{booktabs}
\usepackage{threeparttable}
\usepackage{multirow}
\usepackage{diagbox}
\usepackage[dvipsnames]{xcolor}
\usepackage{booktabs}

\usepackage[ruled,linesnumbered]{algorithm2e}

\definecolor{cvprblue}{rgb}{0.21,0.49,0.74}
\usepackage[pagebackref,breaklinks,colorlinks,citecolor=cvprblue]{hyperref}

\def\paperID{8644} % *** Enter the Paper ID here
\def\confName{CVPR}
\def\confYear{2025}

\title{Infighting in the Dark: Multi-Label Backdoor Attack in Federated Learning}

\makeatletter
\newcommand{\printfnsymbol}[1]{%
  \textsuperscript{\@fnsymbol{#1}}%
}
\makeatother

\author{Ye Li\textsuperscript{\rm 1}, 
Yanchao Zhao\textsuperscript{\rm 1}\thanks{Corresponding authors.}, 
Chengcheng Zhu\textsuperscript{\rm 2}, 
Jiale Zhang\textsuperscript{\rm 2*} \\
\textsuperscript{\rm 1} Nanjing University of Aeronautics and Astronautics \quad
\textsuperscript{\rm 2} Yangzhou University
\\
\tt\small \{miles.li,yczhao\}@nuaa.edu.cn \\
\tt\small chengchengzhu2022@126.com, jialezhang@yzu.edu.cn
}

\begin{document}
\maketitle
\begin{abstract}

% Federated Learning (FL) has been demonstrated to be vulnerable to backdoor attacks. As a decentralized machine learning framework, most research focuses on the Single-Label Backdoor Attack (SBA), where adversaries share the same target but neglect the fact that adversaries may be unaware of each other's existence and hold different targets, i.e., Multi-Label Backdoor Attack (MBA). Unfortunately, directly applying prior work to the MBA would not only be ineffective but also potentially mitigate each other. In this paper, we first investigate the limitations of applying previous work to the MBA. Subsequently, we propose Mirage, a novel multi-label backdoor attack in federated learning (FL), which adversarially adapts the backdoor trigger to ensure that the backdoored sample is processed as clean target samples in the global model. Our key intuition is to establish a connection between the trigger pattern and the target class distribution, allowing different triggers to activate backdoors along clean activation paths of the target class without concerns about potential mitigation. Extensive evaluations comprehensively demonstrate that Mirage outperforms various state-of-the-art attack methods. This work aims to alert researchers and developers to this potential threat and to inspire the design of effective detection methods. Our code will be made available later \href{https://github.com}{Github}.

% Federated Learning, a decentralized machine learning framework, has been demonstrated to be vulnerable to backdoor attacks.
   
Federated Learning (FL), a privacy-preserving decentralized machine learning framework, has been shown to be vulnerable to backdoor attacks. Current research primarily focuses on the Single-Label Backdoor Attack (SBA), wherein adversaries share a consistent target. However, a critical fact is overlooked: adversaries may be non-cooperative, have distinct targets, and operate independently, which exhibits a more practical scenario called Multi-Label Backdoor Attack (MBA). Unfortunately, prior works are ineffective in the MBA scenario since non-cooperative attackers exclude each other. In this work, we conduct an in-depth investigation to uncover the inherent constraints of the exclusion: similar backdoor mappings are constructed for different targets, resulting in conflicts among backdoor functions. To address this limitation, we propose Mirage, the first non-cooperative MBA strategy in FL that allows attackers to inject effective and persistent backdoors into the global model without collusion by constructing in-distribution (ID) backdoor mapping. Specifically, we introduce an adversarial adaptation method to bridge the backdoor features and the target distribution in an ID manner. Additionally, we further leverage a constrained optimization method to ensure the ID mapping survives in the global training dynamics. Extensive evaluations demonstrate that Mirage outperforms various state-of-the-art attacks and bypasses existing defenses, achieving an average ASR greater than 97\% and maintaining over 90\% after 900 rounds. This work aims to alert researchers to this potential threat and inspire the design of effective defense mechanisms. Code has been made \href{https://github.com/NUAA-SmartSensing/Mirage}{open-source}.
\end{abstract}

% Subsequently, we propose Mirage, a novel multi-label backdoor attack in federated learning, which dynamically adapts the backdoor trigger to ensure that the trigger-embedded sample is processed as clean target samples by the backdoored global model. Our key intuition is to establish a connection between the trigger pattern and the target class distribution, allowing different triggers to activate backdoors along clean activation paths of the target class to avoid potential collision. 

% Our code will be made available at \href{https://github.com/NUAA-SmartSensing}{https://github.com/NUAA-SmartSensing}.https://github.com/MMMMMirage/Mirage

\section{Introduction}
\label{Sec: Introduction}

\begin{figure}
    \centering
    \setlength{\abovecaptionskip}{2 mm}
    \includegraphics[width=0.95\linewidth]{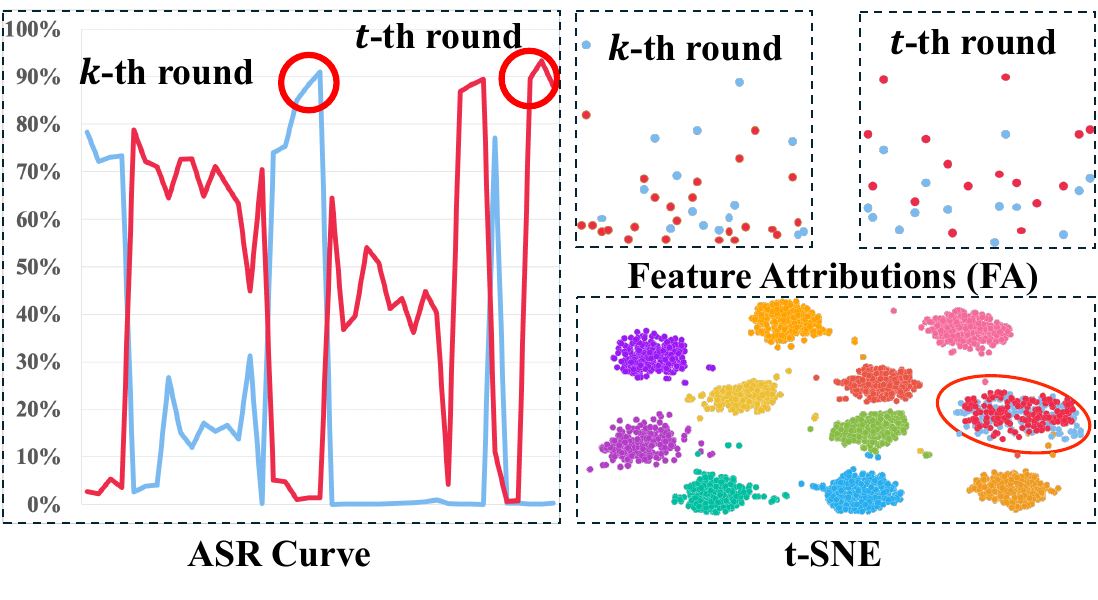}
    \caption{“Blue” and “Red” construct similar OOD mappings (see t-SNE), resulting in a conflict for the neuron weights. Only the dominant attacker can induce the model to output high feature attributions for its backdoor samples (see FA of $k$-th round) and successfully perform the attack. Similar in $t$-th round.}
    \label{Fig: toy_experiment}
    \vspace{-7 mm}
\end{figure}

Recent years have witnessed the flourishing of federated learning, an advanced privacy-preserving distributed learning paradigm \cite{FL_mcmahan_2017, HeteroFair_Li_2024}. FL allows participants to collaboratively build a neural network model without leaking their data. Following its paradigm, a wide range of FL applications have emerged in various safety and privacy-critical application domains, including medical image analysis \cite{Harmofl_jiang_2022, MIA_guan_2024}, face recognition \cite{FaceRecog_niu_2022}, and others \cite{GBoard_hard_2018, QuickType_Apple_2019}.
% and various safety and privacy-critical application domains, including medical image analysis \cite{Harmofl_jiang_2022, MIA_guan_2024}, face recognition \cite{FaceRecog_niu_2022}.

% For backdoor attacks in FL, an attacker injects a backdoor trigger into a subset of their training data, labeling this data with a chosen target label. This poisoned data is then used to train a backdoor model, which is submitted to the FL system, causing the global model to inherit the backdoor.
Despite enhanced privacy, the distributed nature of FL makes them more vulnerable to various adversarial attacks, with \textbf{\textit{backdoor attacks}} posing the greatest threat. For backdoor attacks in FL, an attacker injects a backdoor into the local model, which is then submitted to the FL system, causing the global model to inherit the backdoor after aggregation. Consequently, the global model functions correctly on clean data but misclassifies inputs stamped with the attacker's chosen backdoor trigger as the target class. Numerous studies have been done to improve the persistence and effectiveness of backdoors \cite{A3FL_Zhang_2024, backdoorFL_bagdasaryan_2020, Neurotoxin_zhang_2022, Chameleon_dai_2023}, which pose significant threats to various applications, including computer vision \cite{CV1_wenger_2021}, IoT networks \cite{IoT1_nguyen_2020}, and healthcare systems \cite{healthcare1_jin_2023}.

Recent advancements in backdoor attacks in FL, most, if not all, assume that attackers in training are colluding and sharing the same target class, which can be denoted as Single-Label Backdoor Attacks (SBA). However, situations become more complex in real-world FL applications, especially in large-scale FL scenarios \cite{Fedscale_Lai_2022, FLCMF_Li_2020}, where attackers engage in training individually and compromise the global model for their own purposes, i.e., Multi-Label Backdoor Attack (MBA). Intuitively, non-cooperative adversaries separately attack the global model and do not interfere with each other. Unfortunately, our empirical results indicate that directly applying existing attack methods to multiple adversaries leads to disorganized obstacles, making backdoors not as effective as they are in SBA scenarios. As illustrated in Fig.~\ref{Fig: toy_experiment}, two backdoors cannot be effective simultaneously because attackers construct similar Out-of-Distribution (OOD) backdoor mappings, indicating that their backdoor samples distribute together but outside the distributions of the target classes. This results in attackers excluding one another, with only the dominant attacker capable of embedding a backdoor in the global model, consequently causing the global model to output high feature attributions for its backdoor samples, thereby achieving a high attack success rate (ASR). It further raises a new problem: \textbf{\textit{Can MBA successfully launch without collusion?}}

% As we illustrate in Fig.~\ref{Fig: toy_experiment}, two backdoor features distributed together in feature space represent they construct a similar backdoor mapping in the feature space and the change of ASR curves indicates the shift of its ownership. Specifically, in $t$-th round, “Red” takes the advantage, it outputs relatively high feature attribution values,and correspondingly achieves a high ASR, and the backdoor samples of “Blue” lower its feature attribution values. Things come the same in $k$-th round where “Blue” takes the advantage. This raises a new landscape: \textbf{\textit{Can MBA successfully launch without collusion?}}

To address the aforementioned issue, we first conduct an in-depth investigation to uncover the inherent constraints of the exclusions of SBAs: they create similar backdoor mappings for distinct targets, resulting in conflicts among backdoor functions. Motivated by this observation, we propose Mirage, the first non-cooperative MBA strategy in FL, wherein multiple adversaries individually undermine the global model without collusion. At a high level, instead of coordinating the various OOD mappings, Mirage aims to ensure that the backdoor samples are processed as clean samples by constructing an effective and persistent ID mapping between backdoor features and target distributions. To achieve this goal, we introduce a trigger optimization method based on adversarial adaptation, which adaptively optimizes the trigger to fool a carefully designed OOD sample detector, thereby constructing the ID mapping. Furthermore, considering that the distributions change with global training, we further introduce a constrained optimization method to enhance the persistence of ID mapping by tightening the distribution of backdoor features, allowing the backdoor mapping to live within the global training dynamics. In summary, our contributions are as follows:
\begin{itemize}

    \item \textbf{Uncovering the inherent constraints of extending existing SBAs to MBAs.} To the best of our knowledge, we are the first to investigate and elucidate the inherent constraints of extending existing SBAs to MBAs. We underscore the practical threat posed by such attacks in large-scale FL applications by proposing a multi-label backdoor attack method based on the in-distribution backdoor.

    \item \textbf{Effective ID Mapping Construction.} We propose a novel backdoor optimization method based on adversarial adaptation, which enables attackers to build the bridge between their backdoor and the distribution of their target class, allowing multiple adversaries to simultaneously launch backdoor attacks without interfering with others. 
    
    \item \textbf{Persistent ID Mapping Enhancement.} We additionally introduce a constrained optimization method to enhance the ID mapping by minimizing the similarity of features between backdoored samples and benign samples. This tightens the ID backdoor mapping and ensures its persistence within the global training dynamics.
    
    \item \textbf{Evaluations and Discussions.} Extensive experiments demonstrate that Mirage outperforms SOTA attacks across various settings. Furthermore, we discuss the vulnerability of existing defense methods to MBAs and present potential countermeasures against Mirage, highlighting the urgent demand for tailored defenses.
\end{itemize}

\section{Related Work}
\label{Sec: Related Work}
\subsection{Backdoor Attacks in Federated Learning}

Backdoor attacks aim to compromise the victim model to produce the attacker-desired outputs if and only if the inputs are assigned to a specific trigger pattern. In light of this, a few researchers also start to notice the threat of backdoor attacks on FL systems. Depending on whether the backdoor trigger is optimized, current backdoor attacks fall under the categories of either static-trigger attacks \cite{backdoorFL_bagdasaryan_2020, Neurotoxin_zhang_2022, Chameleon_dai_2023, Darkfed_li_2024, Badfss_zhang_2024} or trigger-optimization attacks \cite{A3FL_Zhang_2024,F3BA_Fang_2023, Advdoor_zhang_2021, CerP_Lyu_2023}. 

The former pre-selected a fixed trigger pattern to poison the local training set and combine them with manually manipulating methods to improve the backdoor effectiveness. For example, \cite{backdoorFL_bagdasaryan_2020, analyzing_bhagoji_2019} aim to replace the global model by uploading a scaled-up poisoned model. In advance, \cite{Neurotoxin_zhang_2022, Towards_Shi_2024, F3BA_Fang_2023} suggest leveraging the unimportant redundant neurons to prevent the backdoor from being erased shortly. Particularly, Dai et al. \cite{Chameleon_dai_2023} introduce contrastive learning from the feature perspective of the backdoor to enhance its durability. The latter optimizes trigger patterns according to different tasks and generally achieves better effectiveness and utility. 3DFed \cite{3DFed_Li_2023} and A3FL \cite{A3FL_Zhang_2024} introduced the backdoor unlearning in an adversarial way to obtain a robust trigger to make the attacks more covert and persistent. DBA \cite{DBA_Xie_2020} and NBA \cite{NBA_nguyen_2024} are the most similar work with us. However, DBA focuses on attacking the same target class, which is also an SBA situation. NBA investigated the non-cooperative multi-label attack in FL but neither revealed the inner reason for the inapplicability of previous works nor proposed an effective method.

\subsection{Backdoor Defenses in FL}

Recently, researchers have noticed the threat of backdoor attacks in FL systems and actively explored various backdoor defense techniques, which can be categorized into either detection \cite{Foolsgold_Fung_2022, RFLBAT_Wang_2022, Backdoorindicator_Li_2024} or mitigation methods \cite{ADFL_CC_2023, FLAME_Nguyen_2022, multikrum_Peva_2017, BadCleaner_JLZ_2024, FLPurifier_JLZ_2024}.

The detection-based methods aim to utilize anomaly detection techniques to determine whether a model is backdoored. Foolsgold \cite{Foolsgold_Fung_2022} assigns lower aggregation weights to updates with high pairwise cosine similarities, thereby mitigating the impact of backdoor updates. RFLBAT \cite{RFLBAT_Wang_2022} discriminates the malicious model according to the difference between poisoned updates and clean updates in low-dimensional projection space. In advance, Li et al. \cite{Backdoorindicator_Li_2024} proposed BackdoorIndicator (referred to as "Indicator" hereafter), which detects potentially poisoned models based on the OOD properties of backdoor samples. Mitigation-based methods explored how to purify the backdoor model by eliminating backdoor triggers. MultiKrum \cite{multikrum_Peva_2017} is a byzantine-robust aggregation protocol that aggregates the global model of having the smallest $\mathcal{L}_2$ distance. FLAME \cite{FLAME_Nguyen_2022} adapts the weak DP method by noise boundary proof and a dynamic clipping bound, which is shown to alleviate the backdoor attack while still retaining a high main task accuracy.

\section{Attack intuitions}
In this section, we investigate the inherent constraints that affect the effectiveness of backdoors when extending previous works to MBA scenarios and further provide the attack intuitions based on this investigation.
% \subsection{Adversaries' capability and knowledge} In this paper, we consider the federated learning system with running image classification tasks. We assume that the adversaries are FL participants, which have full access to their local dataset and training process. The adversaries also have white-box access to the global model weights and predictions. However, adversaries have no access to the dataset of other clients or adversaries. Different from previous work, in which multiple attacker target labels are consistent, we define that the attacker has no information about the others and is unaware of their existence. 

% \subsection{Adversaries' goal}
% The adversaries aim to plant a backdoor into the global model to make the FL global model misclassify trigger-embedded data while leaving other tasks uninfluenced.

\subsection{Inherent constraints of SBA methods}

Recall that, from the neuron activation perspective, the backdoor function in a model is highly related to the specific activation paths across neurons. This results in the model being sensitive to triggered inputs and producing anomalously high feature attribution values \cite{BadNets_gu_2019, DRUPE_tao_2024, Towards_Shi_2024}. Correspondingly, such high values result in backdoor samples distributed out of the target clean distribution in feature space, i.e. the backdoor function constructs the OOD backdoor mapping \cite{Backdoorindicator_Li_2024}. In MBA scenarios, attackers may adopt similar strategies to construct the activation paths, such as using redundant neurons, which leads to the construction of similar OOD mappings and competition for neuron weights. Specifically, different attackers have varying requirements for the output values of specific neurons within the pathway. However, the dominant attacker controls most neurons, ensuring they produce the desired values, which results in others failing to achieve their backdoor function. As shown in Fig.~\ref{Fig: 2_a}, two attackers aim to construct persistent and effective OOD backdoor mappings by targeting redundant neurons. Such a strategy causes them to leverage overlapped activation paths to achieve different functions, which further results in infighting among adversaries. This drives us to find a solution that allows attackers to select different activation paths non-cooperatively.

\subsection{Attack intuitions and challenges}
\label{Subsec: Attack intuition and challenges}
% 现有后门攻击的内在逻辑：建立OOD映射
% 推导出可能的方法1：在不感知其他攻击者的情况下，构建OOD映射；分析缺点：会留下更加明显的OOD痕迹，会被现有基于OOD检测的防御方法检测

% 进一步推导出可能得方法2： In-D ， 解决In-D的挑战：
% 挑战1： 如何保障ID -> ID对抗优化
% 挑战2： 如何增强 -> 端到端优化

\begin{figure}
    \centering
    \setlength{\abovecaptionskip}{2 mm} 
    \begin{subfigure}[t]{0.145\textwidth}
        \centering
        \fbox{\includegraphics[width=\textwidth]{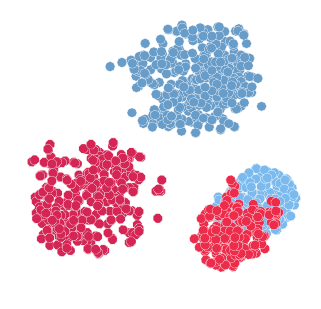}}
        \caption{SBA's mapping.}
        \label{Fig: 2_a}
    \end{subfigure}
        \hfill
    \begin{subfigure}[t]{0.145\textwidth}
        \centering
        \fbox{\includegraphics[width=\textwidth]{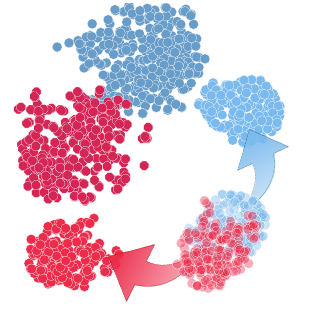}}
        \caption{Separate mapping.}
        \label{Fig: 2_b}
    \end{subfigure}
        \hfill
    \begin{subfigure}[t]{0.145\textwidth}
        \centering
        \fbox{\includegraphics[width=\textwidth]{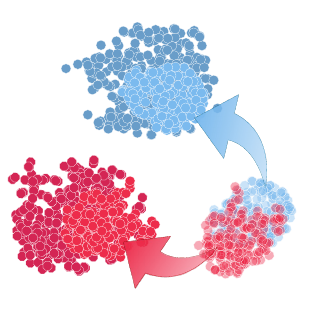}}
        \caption{ID mapping.}
        \label{Fig: 2_c}
    \end{subfigure}
    \vspace{-1.5 mm}
        \hfill
\caption{Illustrations of different mapping strategy.}
\label{Fig: Illustrations}
\vspace{-8 mm}
\end{figure}

% \begin{figure}
%     \centering
%     \includegraphics[width=0.7\linewidth]{Figs/attack_intuition.pdf}
%     \caption{Backdoor Attacks in FL.}
%     \label{Fig: BAinFL}
%     \vspace{-5mm}
% \end{figure}

% It is except from selecting the same neurons. To overcome the aforementioned issue, an intuitive solution to alleviate the infighting is to prevent multiple attackers from using the same neurons which is relative

% . Specifically, A3FL introduce machine unlearning to address the A3FL can address this problem according to such intuition which aims to enhance the backdoor durability by combining machine unlearning  and adversarial training. This is because the unlearning indirectly simulates similar adversarial behavior and targets the use of adversarial training to enhance individual OOD mapping. 

To alleviate the exclusion, it is critical to design a mapping strategy that ensures different attackers can independent mappings with no intersections. Achieving this is challenging because attackers share no information with others and are unaware of each other's existence. 
Surprisingly, our empirical investigation reveals that A3FL can fulfill the aforementioned requirement by combining machine unlearning \cite{BackdoorUnlearning_zeng_2021, NeuralCleanse_wang_2019}, which indirectly stimulates the eliminations performed by other attackers. As shown in Fig~\ref{Fig: 2_b}, the unlearning method adversarially optimizes the trigger to construct a distinct OOD backdoor mapping that differs from other attackers. Although this method mitigates the exclusions, it enhances the OOD characteristics of the backdoor functions, consequently introducing more challenges in bypassing the OOD-based defense methods, such as Indicator \cite{Backdoorindicator_Li_2024}. The OOD-based defense method leads us to a counterintuitive thought: \textbf{\textit{Can a backdoor be triggered via a clean activation path?}} In other words, construct the ID mapping for the backdoor features and target distributions (present in Fig~\ref{Fig: 2_c}). Therefore, once the adversaries construct the ID backdoor mapping to the target class, they naturally alleviate potential exclusions, even without collusion, since additional measures are no longer needed to circumvent the backdoor activation paths of other attackers. However, constructing the ID backdoor mapping remains challenging due to the following issues:

\textit{\textbf{Challenge 1: How to construct the ID mapping?}}

While backdoor attacks have emerged as a prominent topic in FL, to the best of our knowledge, no prior studies investigated the ID backdoor within FL. Intuitively, the backdoor sample should be processed as a clean sample following the clean activation path, meaning it should share similar feature outputs. One promising way is leveraging feature alignment. However, this may disrupt the relationships among clean classes, making it challenging for attackers to maintain the ID characteristics. Therefore, there is a pressing need to devise a novel method that ensures the effective construction of ID mapping.

\textit{\textbf{Challenge 2: How to maintain the persistence of ID mapping?}}

One key challenge for backdoor attacks in FL is that the continued dynamic clean updates may break the well-constructed backdoor mappings. Previous research focuses on leveraging neurons that are updated less frequently to build the neural activation path, which cannot be applied in ID mapping since our objective is to construct the bridge between the trigger pattern and the clean target activation path. Therefore, these maintenance requirements for ID mapping further exacerbate the challenges.

\section{Methodology}
\label{Sec: Methodology}
To formulate the attack scenario, we first introduce the threat model. Driven by attack intuitions, we propose two methods that enable attackers to construct an efficient and persistent ID mapping without cooperation.
\subsection{Threat model} 
\noindent\textbf{Adversaries’ Capabilities and knowledge}. We consider a FL system running image classification tasks and assume that adversaries are FL participants, which enables attackers to control their own local datasets and training processes. Note that adversaries do not have access to any information regarding benign clients or other adversaries, including but not limited to datasets, models, and target classes.
% , which means they have the full control over the training process and have the access to its local datasets. Besides, attackers also have the access to the global model if they are selected to the have full access to their local datasets and training processes. The adversaries also have access to the global model weights and predictions. However, adversaries do not have access to any information of benign clients or other adversaries, including their datasets, models, target classes and others.
% the dataset and models of other clients or adversaries. Different from previous work, in which multiple attacker target labels are consistent, we define that the attacker has no information about the other attackers.

\noindent\textbf{Adversaries' goal.} Adversaries aim to plant a backdoor into the global model that makes it misclassify backdoor data to a specific class while leaving other tasks unaffected.

\begin{figure*}
    \centering
    \setlength{\abovecaptionskip}{2 mm}    \includegraphics[width=0.9\linewidth]{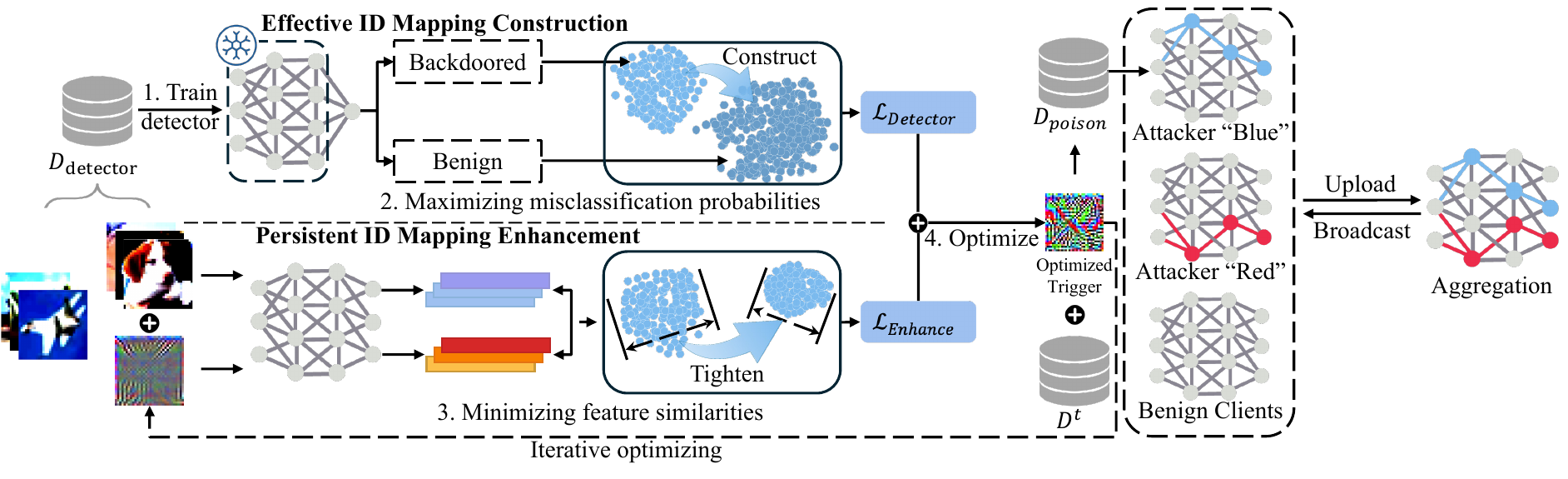}
    \caption{The workflow of Mirage. \textbf{Step 1}: Train an OOD sample detector $\theta_{Detector}$. \textbf{Step 2}: Construct ID mapping by maximizing misclassification probabilities of backdoor samples on the detector. \textbf{Step 3}: Tighten backdoor distribution by minimizing the feature similarities between backdoor samples and benign samples. \textbf{Step 4}: Optimize the trigger by minimizing $L_{Enhance}$ and $L_{detector}$. }
    \label{Fig: workflow}
    \vspace{-4 mm} 
\end{figure*}
% 本章中，我们详细描述我们所提出的框架，其通过建立后门特征到干净分布之间的干净映射来完成无串通情况下的MBA.
%在本章节，我们展示了所提出框架的细节，其允许攻击者构建从后门特征到目标类别分布的分布内映射，以此使得攻击者可以无惧其他攻击者的潜在抵消
\subsection{Overview of Mirage}
Following the attack intuition, our objective is to construct an ID mapping to bridge the backdoor features and distribution of the target class, thereby freeing attackers from the exclusion of using similar backdoor activation paths. This presents two significant challenges: how to construct a qualified ID mapping and how to maintain it. To address these challenges, we propose Mirage, the first multi-label backdoor attack in FL, which enables attackers to non-cooperatively compromise the global model by constructing an ID mapping between the backdoor features and target distributions. As illustrated in Fig.~\ref{Fig: workflow}, Mirage introduces two components to construct an effective and persistent ID backdoor mapping. On the one hand, we propose a backdoor optimization method based on adversarial adaptation, which constructs the ID mapping by maximizing the misclassification probability of a well-designed backdoor sample detector. On the other hand, we deploy a constrained optimization strategy to tighten the backdoor distribution, ensuring its survival in the global training dynamics. Next, we provide details of each component.
\subsection{Effective ID Mapping Construction}
\label{Subsec: ID mapping construction}
% 如何解决挑战1：构建ID mapping
To address the first challenge, we aim to adversarially optimize the trigger pattern to make it able to construct the ID mapping. In order to optimize the trigger, we employ an OOD sample detector to determine whether a backdoored sample is an OOD sample and further optimize the trigger pattern to fool the detector. This introduces another problem: How can an attacker obtain an accurate detector? Previous research suggests training a new model as the detector \cite{Advdoor_zhang_2021}. However, retraining the detector with each trigger update incurs significant computational overhead, and the local samples limit its effectiveness. These shortcomings prompt us to seek another solution to obtain a detector that is both effective and lightweight for updates. Recall that the FL training process allows clients to obtain the latest global model, which contains a well-trained feature extractor. Consequently, we consider leveraging the feature extractor of the global model as the primary component and training a binary classifier to distinguish the OOD samples. Such an approach is a common practice in transfer learning, consumes only a few computing resources, and requires little data.

% Recall the tradition backdoor attacks again, they focus on building a shortcuts path towards the target class, produces an OOD mapping to the clean samples, which means that the global model,especially the feature extractor, process the clean sample and backdoored sample in different way. Thus, a directly solution for detector is that we can leverage the feature extractor of global model as the major component of detector and train a binary-classifier to discriminate the clean samples.
In order to obtain an accurate detector, we first prepare the dataset for training. The local dataset for each attacker can be divided into $D_{cln}=\{(x,y)\in D_{att})|+\}$ and $D_{bd} = \{(x\oplus\delta,y)\in D_{att})|-\}$, where $+$ indicates the ground truth label of a sample is $c$, and $-$ is the opposite and $\delta$ represents the trigger pattern. Then, the detector dataset $D_{detector}$ ($D_{d}$ for short) is composed as follows:
\begin{equation}
\begin{array}{rr}
   D_{d} = \{(x,-) \mid x \in D_{bd}\} \cup \{(x,+ ) \mid x \in D_{cln}\},
\end{array}
\label{Equ: detector dataset construct}
\end{equation}

% In order to obtain an accurate detector, the dataset is also a necessary precondition. We directly divide the local dataset of an attacker to $D_{clean}=\{(x,y)\in D_{attacker})|y=c\}$, where $c$ is the target class for the attacker. The left samples are making up to $D_{backdoor} = \{(x\oplus\delta,y)\in D_{attacker})|y\neq c\}$ where $\delta$ represent the trigger pattern. Further, the detector dataset $D_{detector}$ is composed as follows:
% \begin{equation}
% \begin{array}{ll}
%    D_{detector} =  \\
%         \{(x,false) \mid x \in D_{backdoor}\} \cup \{(x,true) \mid x \in D_{clean}\}.
% \end{array}
% \label{Equ: detector dataset construct}
%  % \{(x,false) \mid x \in D_{backdoor}\} \cup \{(x,true) \mid x \in D_{clean}\}.
% \end{equation}

To make the ID mapping construction more effective and flexible, we adopt an adversarial training-like method to optimize the trigger pattern. Specifically, $\theta_{detector}$ is trained on the $D_{detector}$ to differentiate between clean and backdoored samples by minimizing the binary cross-entropy (BCE) loss. 

Concurrently, the trigger is optimized to fool the detector by maximizing the binary classification loss. Such a min-max game equips the trigger with the ability to activate the clean activation path of the target class in the global model, naturally avoiding conflicts with other attackers. The min-max process is mathematically described as follows:
\begin{equation}
\begin{array}{c}
\delta=\underset{\delta}{\ \operatorname*{argmin}} \ \mathbb{E}_{(x,y)\sim D_{detector}}[\mathcal{L}(x,+,\theta)|y=-]
\\
    s.t.\ \theta = \ \underset{\theta}{\operatorname*{argmin}}\  \mathbb{E}_{(x,y)\sim D_{detector}}[\mathcal{L}(x,y,\theta)]
\end{array}
\end{equation}
where $\theta$ is the detector's parameters, with its feature extractor frozen and parameterized by the weights of the global model, and $\mathcal{L}$ is the BCE loss. For better implementation, we replace the maximization problem with minimizing the BCE loss of misclassification for the backdoored samples.

% Specifically, we aim to train a OOD detector $\theta_{detector}$ to discriminate whether a sample is backdoored, and further adversarial optimize the trigger to fool the detector. Previous research proposed to train an MLP as the discriminator, which is not appropriate here, since iterative optimize the trigger pattern request the detector have sufficient data samples which is not realistic in FL. 

% we adopt an adversarial training-like method to optimize the trigger pattern to build the bridge between the backdoored feature and target clean distribution.
% % 利用对抗训练的思想，优化trigger，具体是构建一个OOD探测器，通过不断对trigger优化，来使得trigger可以表现出目标类的干净特征
% Specifically, we aim to train an OOD detector $\theta_{detector}$ to discriminate whether a sample is backdoored, and further adversarial optimize the trigger pattern to fool the detector. To achieve this, a binary detector should be trained and updated with trigger.

% % 如何构建detector - 数据集构建

% After that, we 
% % 如何构建训练集
% % 对抗优化的目标
% In this subsection, we detail the design of the OOD detector $\theta_{detector}$ and how can we optimize the trigger pattern to ensure the backdoored sample that are in-distribution.

\subsection{Persistent ID Mapping Enhancement}
\label{Subsec: ID mapping enhancement}
% 为什么要做增强： 因为目前的受限于攻击者本地的数据，训练出的detector会有欠缺，也就是不能表现出较强的后门特征

% 这是因为detector受限于本地样本不足以及目标类样本的表征之间存在一定差异，导致先前生成的trigger无法使所有被后门的样本处于分部内，由于我们采用激活干净神经通路的方式植入后门，这些样本无法有效的激活目标通路，造成了攻击的成功率不及预期。在以往的工作中同样存在这样的情况，一些工作提出利用GAN或者生成模型来丰富本地数据，或者使用KDE方法估计目标分布，然后通过数据增强来增加本地样本，但是这种方式极大的增加了计算开销。
Although using an OOD detector and an adversarial strategy restricts the trigger pattern $\delta$ to reveal the ID feature, our empirical results indicate that the backdoored samples still do not meet our expectations. This is because the adversarial strategy does not constrain the tightness of backdoor mapping, which causes the margin-distributed backdoored samples not in-distribution anymore with the dynamic changes of training and failing to activate the target activation path. One promising solution is to leverage distribution estimation methods, like KDE \cite{KDE_sheather_1991}, to obtain the actual distribution of the target class and use data augmentation to generate samples to improve the effectiveness of the detector. Although it has been proven effective, it requires countless computing resources. Therefore, we introduce a constrained optimization method to enhance the ID mapping by tightening the distribution of backdoored samples. Since the adversarial strategy successfully constructs the ID mapping, we aim to “push” the backdoor distribution to make it tighter. Specifically, we minimize the similarities between clean samples $x$ and their corresponding backdoored samples $x \oplus \delta$, allowing the backdoored sample to deviate as much as possible from the original distribution. Additionally, we establish the direction by minimizing the loss of the backdoor samples on the global model. Formally, it can be formulated as the following optimization problem:
\begin{equation}
\begin{array}{r}
 \delta = \operatorname{argmin}\ \operatorname{CS}[(\theta_{f}(x),\theta_{f}(\hat{x})]  +  \operatorname{CE}(\hat{x},\hat{y}, \theta)
\end{array}
\end{equation}
where $\theta_{f}$ represents the feature extractor of the global model $\theta$. $\hat{x} = x \oplus \delta, x\in D_{att}$, $\oplus$ is the backdoored operator. $\operatorname{CS(\cdot)}$ and $\operatorname{CE(\cdot)}$ represent the Cosine-similarity and the Cross-entropy loss, separately. Therefore, the backdoor distribution can be tightened into the target distribution and live within the global training dynamics.

% Specifically, we aim to maximize the difference between the clean sample $x$ and its corresponding backdoored sample $x \oplus \delta$, allowing the backdoored sample to deviate as much as possible from the original distribution. Additionally, we maintain the tightness of the backdoor mapping in the target distribution by minimizing the loss of the backdoor samples on the global model, i.e., enhancing their ID mapping in the target distribution. Formally, the ID mapping enhancement can be formulated as the following optimization problem:

We summarized Mirage in Algorithm~\ref{Algo: Mirage}, and the detailed description is provided in Appendix ~\ref{Appendix: Algorithm Outline}.

\begin{algorithm}[h]

\caption{Mirage at $i$-th attacker.}

\label{Algo: Mirage}
\KwData{$\theta_t,\theta_{detector},\theta_{fe},\mathcal{D}_i,\delta,\eta, c, E$.}

$\theta_t^{i} = \theta_t $\;
\textit{/* Trigger adversarial adaption*/}

\For{epoch $e=1$, \dots, $E$}{
    Generate $D_{detector}$ by Eq~\ref{Equ: detector dataset construct}\;
    $\theta_{detector}= \underset{\theta}{\operatorname*{argmin}}\  \mathbb{E}_{(x,y)\sim D_{detector}}[\mathcal{L}(x,y,\theta_{detector})]$\;

    \For{batch $\mathcal{B} \in D_i$}{
    $\mathcal{L}_{detector} = \frac{1}{|\mathcal{B}|}\sum_{(x,y)\in\mathcal{B}}\mathcal{L}_{BCE}(x\oplus\delta, + ,\theta_{detector})$\;

    $\mathcal{L}_{Enhance} = \frac{1}{|\mathcal{B}|}\sum_{(x,y)\in\mathcal{B}}\mathcal{L}_{CE}(x\oplus\delta, \hat{y},\theta_t^{i})\hfill$ \\ 
    $ \hfill+  \mathcal{L}_{CS}((\theta_{fe}(x),\theta_{fe}(x \oplus \delta));$
    
    $\delta\leftarrow \delta - \eta \cdot \nabla_{\delta}(\mathcal{L}_{detector} + \mathcal{L}_{Enhance});$
    }
}
\textit{/* Local training*/}

Get the poisoned local dataset $\mathcal{D}_i^{poison}$ with $\delta$\;

Local optimize $\theta_t^{\prime}$ on $\mathcal{D}_i^{poison}$\;

$\boldsymbol{\Delta}_i^{t+1}=\theta_t-\theta_t^\prime$\;

Upload $\boldsymbol{\Delta}_i^{t+1}$ to the server\;

\end{algorithm}
\section{Experiments}
\label{Sec: Experimental Evaluation}
In this section, we provide extensive experimental results to demonstrate the effectiveness of Mirage by comparing its performance with several SOTA backdoor attack methods from multiple angles. Our experiments run on a FL system performing image classification tasks, using an NVIDIA GeForce RTX 4090 GPU equipped with 24GB of memory. The detailed setups are listed as follows:

\textbf{Dataset:} We conduct experiments on three widely used public real-world datasets, CIFAR-10, CIFAR-100 \cite{Cifar_Krizhevsky_2009} and GTSRB \cite{GTSRB_houben_2013}. Details are provided in Appendix~\ref{Appendix: datasets}.

\textbf{Federated learning setup:}
By default, we set the number of FL clients $N=100$. The server randomly selects $M = 10$ clients at each communication round to contribute to the global model. The default global model architecture is ResNet-18 \cite{Resnet_He_2016}, which is a classic and effective model for image classification. Based on previous research, we randomly split the dataset among clients non-IID using Dirichlet sampling \cite{Dirichlet_hsu_2019} with a hyperparameter $\alpha=1$. For each selected benign client, they train the local model for two local epochs using an SGD optimizer with a learning rate of 0.01. The batch size is 64 for CIFAR-10 and CIFAR-100 and 32 for GTSRB. The entire training process continues for 2100 communication rounds.

\textbf{Attack and defense setup:} We compare the effectiveness of Mirage with SOTA methods, including A3FL \cite{A3FL_Zhang_2024}, Chameleon \cite{Chameleon_dai_2023}, Neurotoxin \cite{Neurotoxin_zhang_2022}, PGD \cite{PGD_Wang_2020}, and Vanilla \cite{backdoorFL_bagdasaryan_2020}. We further extend the aforementioned methods to MBA scenario, wherein attackers independently execute their attacks. Unless otherwise specified, we set $N=3$, with the target classes being 0, 1, and 2. In line with A3FL, attackers of Mirage optimize their triggers by PGD and poison 12.5\% of $D_{att}$. Blend \cite{Blend_chen_2017} is utilized as a backdoor trigger across all attacks. The attack window starts from the 2000-th round and continues for 100 rounds, so attackers are forced to wait for selection and cannot launch a continuous attack. 
% Attackers start compromising the global model after 2000-th round and continues for 100 rounds. Note that we consider a more realistic attack scenario in which the server randomly selects clients to contribute to the global model, which means that the attacker is forced to wait for selection and cannot launch a continuous attack. 

We also evaluate those attacks against six SOTA defenses: Multi-Krum \cite{multikrum_Peva_2017}, DeepSight \cite{DeepSight_Rieger_2022}, Foolsgold \cite{Foolsgold_Fung_2022}, RFLBAT \cite{RFLBAT_Wang_2022}, FLAME \cite{FLAME_Nguyen_2022} and BakcdoorIndicator \cite{Backdoorindicator_Li_2024}. For those attacks and defense methods, we follow the original implementations. Note that we selected the random noise dataset as the source of the indicator dataset.

\subsection{Evaluation metrics}
Following previous works, we comprehensively evaluate Mirage using accuracy (Acc), attack success rate (ASR), and lifespan. Specifically, Acc reflects the attack stealthiness, which indicates the performance of the backdoor model on the main tasks. ASR represents the ratio of backdoored samples misclassified as the labels specified by the attackers. Note that we report the mean ASR of three attackers on the global model at the end of FL. To evaluate the persistence of our attack, we introduce lifespan, denoted as the period that starts at the end of the attack window and ends when the ASR decreases to below a chosen threshold.

\begin{table*}
\resizebox{\textwidth}{!}{%
\begin{tabular}{c|c|ccc|ccc|ccc|ccc|ccc|ccc}
\toprule
\multirow{2}{*}{Dataset}   & \multirow{2}{*}{Defense} & \multicolumn{3}{c|}{Vanilla}                                & \multicolumn{3}{c|}{PGD}                                    & \multicolumn{3}{c|}{Neurotoxin}                              & \multicolumn{3}{c|}{Chameleon}                              & \multicolumn{3}{c|}{A3FL}                                   & \multicolumn{3}{c}{Mirage}                                   \\ \cmidrule{3-20} 
                           &                         & \multicolumn{1}{c|}{Acc ($\uparrow$)} & \multicolumn{1}{c|}{ASR ($\uparrow$)} & GAP ($\downarrow$)   & \multicolumn{1}{c|}{Acc ($\uparrow$)} & \multicolumn{1}{c|}{ASR ($\uparrow$)} & GAP ($\downarrow$)   & \multicolumn{1}{c|}{Acc ($\uparrow$)} & \multicolumn{1}{c|}{ASR ($\uparrow$)} & GAP ($\downarrow$)    & \multicolumn{1}{c|}{Acc ($\uparrow$)} & \multicolumn{1}{c|}{ASR ($\uparrow$)} & GAP ($\downarrow$)   & \multicolumn{1}{c|}{Acc ($\uparrow$)} & \multicolumn{1}{c|}{ASR ($\uparrow$)} & GAP ($\downarrow$)   & \multicolumn{1}{c|}{Acc ($\uparrow$)} & \multicolumn{1}{c|}{ASR ($\uparrow$)} & GAP ($\downarrow$)  \\ \midrule
\multirow{7}{*}{CIFAR-10}  & nodefense               & 91.76                    & 31.88*                   & 87.68 & 91.52                    & 31.25*                   & 92.78 & 92.03                    & 52.99*                   & 90.29  & \textbf{92.35}           & 39.65*                   & 79.98 & 91.73                    & 99.52*                   & 0.23  & 92.16                    & \textbf{99.54*}          & 0.27 \\
                           & deepsight               & 91.44                    & 31.71*                   & 88.29 & 90.81                    & 31.083*                  & 92.63 & 92.31                    & 39.24*                   & 83.50  & \textbf{92.32}           & 36.66*                   & 79.66 & 92.17                    & 96.56*                   & 7.74  & 91.28                    & \textbf{97.39*}          & 3.82 \\
                           & foolsgold               & 88.79                    & 32.50*                   & 89.50 & 90.71                    & 31.67*                   & 93.96 & 91.04                    & 80.91*                   & 21.83  & 90.66                    & 45.46                    & 67.25 & 89.00                    & 95.30*                   & 12.02 & \textbf{91.32}           & \textbf{97.35*}          & 2.51 \\
                           & Indicator               & 85.08                    & 28.76                    & 85.32 & 85.83                    & 20.60                    & 59.50 & 85.13                    & 29.98                    & 60.32* & 86.13                    & 8.21                     & 5.13  & 85.75                    & 70.33*                   & 88.45 & \textbf{91.10}           & \textbf{93.46*}          & 3.90 \\
                           & multikrum               & 92.36                    & 1.59                     & 4.56  & 92.39                    & 1.18                     & 3.25  & \textbf{92.64}           & 1.89                     & 5.18   & 92.46                    & 1.90                     & 5.25  & 92.18                    & 78.15*                   & 64.41 & 92.10                    & \textbf{92.30*}          & 3.12 \\
                           & rflbat                  & 86.83                    & 30.55                    & 88.34 & \textbf{88.28}           & 33.27*                   & 99.79 & 87.22                    & 33.27*                   & 73.12  & 86.97                    & 37.15                    & 86.42 & 85.46                    & 87.83*                   & 10.21 & 85.74                    & \textbf{97.29*}          & 2.11 \\
                           & Flame                   & \textbf{92.56}           & 1.83                     & 5.26  & 92.51                    & 3.39                     & 9.90  & 92.02                    & 40.97*                   & 94.04  & 92.28                    & 1.38                     & 3.71  & 92.25                    & 79.40*                   & 60.23 & 92.36                    & \textbf{95.9*}           & 2.42 \\ \midrule
\multirow{7}{*}{CIFAR-100} & nodefense               & 69.40                    & 32.62*                   & 94.31 & 70.53                    & 32.28*                   & 93.89 & 71.07                    & 32.62*                   & 97.05  & \textbf{71.66}           & 24.07                    & 27.50 & 71.26                    & 98.67*                   & 2.47  & 71.65                    & \textbf{99.05*}          & 0.77 \\
                           & deepsight               & 70.02                    & 32.86*                   & 89.02 & 70.54                    & 32.23*                   & 96.32 & 70.53                    & 33.14*                   & 93.28  & 71.30                    & 24.06                    & 26.17 & \textbf{71.68}           & 92.05*                   & 18.44 & 71.56                    & \textbf{98.67*}          & 2.43 \\
                           & foolsgold               & 70.02                    & 33.30                    & 89.81 & 70.15                    & 32.65*                   & 97.34 & 71.07                    & 32.63*                   & 97.05  & 71.60                    & 24.20                    & 27.11 & 71.78                    & 98.71*                   & 1.64  & \textbf{71.94}           & \textbf{99.18*}          & 1.00 \\
                           & Indicator               & 67.73                    & 6.80                     & 20.12 & 63.85                    & 5.88                     & 17.28 & 60.98                    & 11.89                    & 32.34  & 61.35                    & 6.07                     & 13.43 & 65.35                    & 37.13*                   & 92.72 & \textbf{68.22}           & \textbf{99.8*}           & 0.10 \\
                           & multikrum               & 71.81                    & 0.38                     & 0.77  & 71.14                    & 0.31                     & 0.59  & 71.63                    & 0.36                     & 0.66   & \textbf{71.85}           & 0.21                     & 0.38  & 71.76                    & 91.28*                   & 18.55 & \textbf{71.86}           & \textbf{93.57*}          & 5.14 \\
                           & rflbat                  & 62.36                    & 0.30                     & 0.41  & 63.10                    & 0.60                     & 0.12  & 62.15                    & 0.56                     & 1.00   & 61.49                    & 18.01                    & 27.31 & 61.35                    & 94.34*                   & 16.56 & \textbf{63.19}           & \textbf{95.68*}          & 6.99 \\
                           & Flame                   & 71.75                    & 0.21                     & 0.28  & \textbf{72.09}           & 0.33                     & 0.62  & 72.02                    & 0.27                     & 0.45   & 72.10                    & 0.30                     & 0.52  & 71.22                    & 91.48*                   & 19.83 & 71.82                    & \textbf{94.22*}          & 3.16 \\ \midrule
\multirow{7}{*}{GTSRB}     & nodefense               & 96.10                    & 33.32*                   & 98.24 & 96.73                    & 33.10*                   & 98.80 & 95.93                    & 38.39*                   & 95.79  & 96.08                    & 45.62*                   & 87.06 & 96.65                    & 99.63*                   & 0.49  & \textbf{96.97}           & \textbf{99.73*}          & 0.48 \\
                           & deepsight               & 95.95                    & 33.8*                    & 97.44 & 95.97                    & 33.36*                   & 98.35 & 96.12                    & 33.06*                   & 99.17  & 96.14                    & 50.14                    & 76.71 & 95.95                    & 94.78*                   & 12.64 & \textbf{96.20}           & \textbf{96.76*}          & 2.53 \\
                           & foolsgold               & 96.08                    & 33.47*                   & 97.92 & 96.94                    & 33.06*                   & 98.75 & 95.88                    & 37.57*                   & 96.36  & 96.04                    & 51.91                    & 78.06 & 96.62                    & 97.68*                   & 6.95  & \textbf{96.95}           & \textbf{99.88*}          & 0.27 \\
                           & Indicator               & 91.57                    & 50.92                    & 91.53 & 89.45                    & 27.27                    & 78.14 & 89.08                    & 17.09                    & 50.97  & 92.91                    & 2.76                     & 5.38  & 85.51                    & 85.05*                   & 25.26 & \textbf{95.12}           & \textbf{99.73*}          & 0.70 \\
                           & multikrum               & 96.24                    & 4.14                     & 12.20 & \textbf{96.97}           & 2.74                     & 7.72  & 96.12                    & 4.84                     & 14.32  & 96.13                    & 4.72                     & 13.97 & 96.62                    & 98.18*                   & 4.16  & 96.92                    & \textbf{98.99*}          & 1.73 \\
                           & rflbat                  & 94.40                    & 56.61*                   & 90.62 & 89.44                    & 33.26*                   & 99.72 & 94.53                    & 33.88*                   & 97.80  & 94.75                    & 54.39*                   & 83.04 & \textbf{95.23}           & 97.86*                   & 6.42  & 94.21                    & \textbf{99.97*}          & 0.08 \\
                           & Flame                   & 95.88                    & 6.19                     & 18.45 & \textbf{96.97}           & 2.74                     & 7.72  & 95.92                    & 5.15                     & 15.32  & 95.92                    & 5.95                     & 17.71 & 96.52                    & 99.31*                   & 1.52  & 96.92                    & \textbf{99.32*}          & 1.05 \\ \bottomrule
\end{tabular}%
}
\caption{Performance of Mirage compared with baseline attack methods against the SOTA defense methods. The Acc higher represents better, and so does the ASR. “GAP” represents the difference between the highest and the lowest ASR, and attack methods should maintain their GAP close to zero. An “*” after ASR indicates that at least one of the three attackers achieved over 90\% ASR.}
\label{Tab: acc_asr}
\vspace{-3 mm}
\end{table*}

\subsection{Attack performance}
\label{Subsection: Attack performance}
% Except for the Acc and ASR, we also report the GAP of the three attackers' ASR, which is the difference between the highest and lowest ASR among the three attackers. Note that a successful attack should achieve a high ASR and maintain the GAP close to zero.

First, we evaluate the attack performance of Mirage across various datasets and compare it to different SOTA attack methods. The results are summarized in Table~\ref{Tab: acc_asr}, and the best results are highlighted in Bold. The experiment shows that Mirage achieves remarkable results across three tasks and various SOTA defense methods compared to the baseline methods. In terms of accuracy, Mirage achieves the best accuracy in over half of the tasks and obtains suboptimal results for the remaining tasks, losing to the best ones by an average of only 0.498\%. Regarding effectiveness, Mirage achieves the best ASRs across all the tasks, with the gap among attackers being acceptable (the maximum gap is 6.99\%, and the average gap is 2.19\%). Those results indicate that the design of constructing the ID mapping between backdoor features and target class allows attackers to independently attack the FL model without concerns about the potential exclusion, resulting in a global model that excels in both main and multiple backdoor tasks. 

Furthermore, A3FL is a relatively competitive method that leverages unlearning to simulate the potential exclusion and optimize the trigger to construct robust activation paths independent of other attackers. However, as described in Sec.~\ref{Subsec: Attack intuition and challenges}, such an approach may enhance its OOD mapping concerning the benign distribution and increase the risk of being detected by OOD-based detection methods, such as Indicator \cite{Backdoorindicator_Li_2024}. The results of A3FL with the defense of Indicator support our analysis. Besides, the Indicator does not perform as expected, we will discuss it in Section~\ref{Subsec: Potential defense methods}.

\subsection{Lifespan}
\label{Subsec: Lifespan}

\begin{figure}
    \centering
    \setlength{\abovecaptionskip}{1 mm}    \includegraphics[width=1\linewidth]{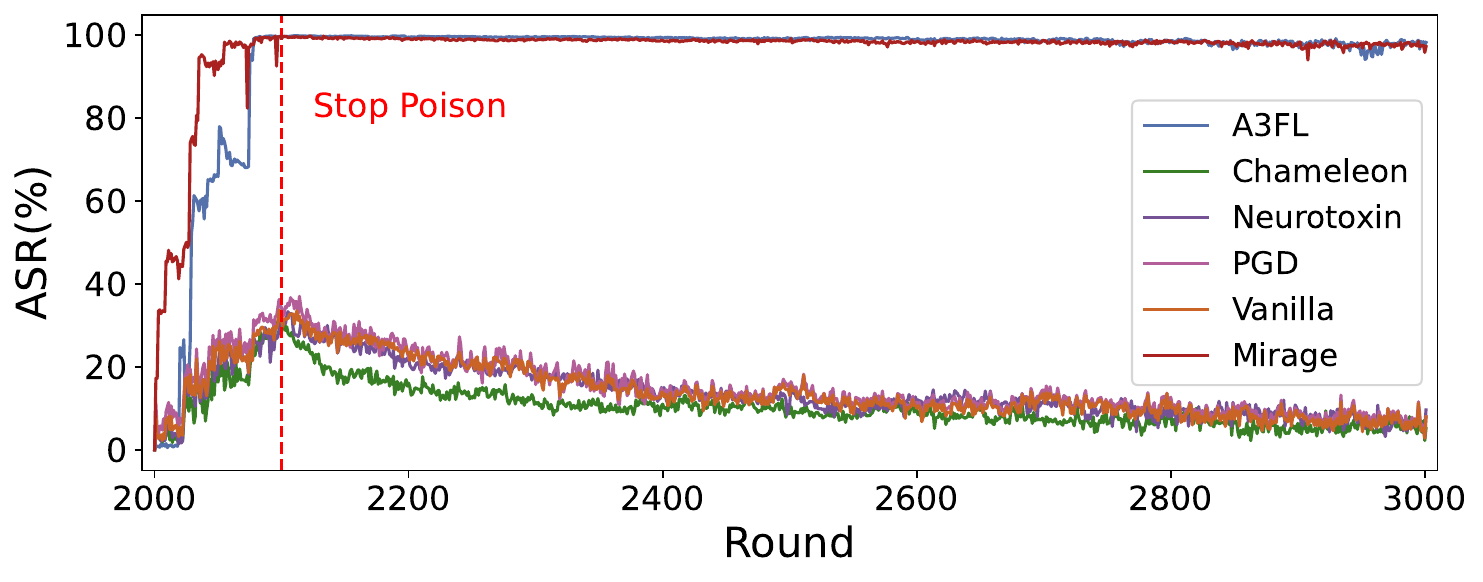}
    \vspace{-3 mm}
    \caption{Persistent evaluations.}
    \label{Fig: Persistent}
    \vspace{-8 mm}
\end{figure}
Secondly, we evaluate the durability of attacks by comparing their lifespans. We extend the FL training process from 2100 to 3000 rounds while keeping the attack window to the first 100 rounds. Fig.~\ref{Fig: Persistent} illustrates the attack success rate without any defense method. Mirage achieves similar results with A3FL, where the ASRs exceed 90\% at the end, indicating a lifespan of over 900 rounds. In contrast, the ASRs of all other baseline attacks are constrained by potential exclusion between attackers, as described before, making them ineffective and falling rapidly. These experimental results suggest that Mirage can achieve comparable durability to A3FL, which focuses on enhancing the persistence of backdoors in FL.

\subsection{The impact of different models}

In this section, we evaluate Mirage across different model architectures. Following previous research, we select widely used image classification model architectures: ResNet18 and ResNet34 \cite{Resnet_He_2016}, VGG11 and VGG19 \cite{VGG_Simonyan_2015}, and MobileNet-V2 \cite{Mobilenetv2_sandler_2018}. We set the parameters to default and only change the model architectures.  As shown in Tab.~\ref{Tab: different models}, the attack performance of Mirage remains high in ASRs. The decrease in MobileNet-V2 is mainly attributed to the low utility of its model performance, which significantly reduces the detector's utility, resulting in sub-optimal ID mapping construction for certain attackers. However, the average ASRs remain acceptable. In conclusion, based on the results presented above, we can conclude that Mirage is not sensitive to specific model structures.

\begin{table}
\centering
\resizebox{0.4\textwidth}{!}{%
\begin{tabular}{cccccc}
\toprule
    & ResNet18 & ResNet34 & VGG11 & VGG19 & MobileNet-V2 \\ \midrule
Acc & 92.16    & 93.89    & 91.14 & 92.07 & 85.38        \\
ASR & 98.8     & 99.63    & 99.21 & 99.82 & 97.65        \\
% MAX & 99.01    & 99.92    & 99.84 & 99.98 & 98.56        \\
% MIN & 98.63    & 99.26    & 98.87 & 99.5  & 95.86        \\ 
\bottomrule

\end{tabular}%
}
\caption{Results on different model architectures.}
\label{Tab: different models} 

\vspace{-3mm}
\end{table}

\begin{table}
\centering
\resizebox{0.35\textwidth}{!}{%
\begin{tabular}{ccccccc}
\toprule
Dataset                   &     & 0.5   & 1     & 5     & 10    & 1000  \\ \midrule
\multirow{2}{*}{CIFAR100} & Acc & 71.46 & 71.65 & 71.68 & 71.50 & 71.81 \\
                          & ASR & 99.59 & 99.05 & 99.16 & 99.26 & 99.21 \\ \midrule
\multirow{2}{*}{GTSRB}    & Acc & 96.96 & 96.97 & 96.62 & 96.57 & 96.70 \\
                          & ASR & 99.54 & 99.73 & 99.47 & 98.26 & 98.86 \\ \midrule
\multirow{2}{*}{CIFAR10}  & Acc & 91.61 & 92.16 & 92.43 & 92.51 & 92.38 \\
                          & ASR & 98.95 & 98.80 & 99.03 & 99.39 & 99.15 \\ \bottomrule
\end{tabular}%
}
\caption{Results on different non-IID settings.}
\vspace{-3 mm}
\label{Tab: different data distribution}
\end{table}

\subsection{The impact of adversary number}
\label{Subsec: different adversary number}

We also study the impact of the number of adversaries, which can be considered as the number of different targets since we assign different targets to the attackers. Table~\ref{Tab: different attacker number} presents the experimental results on CIFAR-10 with varying attack numbers ranging from 1 to 5, evaluating the performance of the global model on the main tasks and different backdoor tasks. Note that other settings follow the default configuration. We observe that the increase in attack numbers has a negligible impact on Mirage, with the accuracy of main tasks maintained above 92\% and the average ASR exceeding 98\%. The experimental results indicate that backdoor injection without collusion can be effectively implemented by constructing ID mapping. We also conducted experiments on different datasets (see Appendix~\ref{Appendix: Different Attacker Numbers}), and the results support the above conclusions.

% \begin{figure}
%     \centering
%     \includegraphics[width=0.7\linewidth]{Figs/different client number.pdf}
%     \caption{Mirage with different attacker number.}
%     \label{Fig: different attacker number}
%     \vspace{-5mm}
% \end{figure}

\begin{table}
\centering
\resizebox{0.35\textwidth}{!}{%
\begin{tabular}{cccccc}
\toprule
Attacker Number  & 1     & 2     & 3     & 4      & 5      \\ \midrule
Acc         & 92.37 & 92.53 & 92.16 & 92.11  & 92.12  \\
ASR         & 99.06 & 99.15 & 98.80 & 98.375 & 98.484 \\ 
GAP & 0.00 & 0.49 & 0.38& 1.42& 1.99\\
% Attacker\_1 & 99.06 & 99.39 & 99.01 & 98.71  & 98.66  \\
% Attacker\_2 & -     & 98.9  & 98.63 & 98.43  & 97.54  \\
% Attacker\_3 & -     & -     & 98.77 & 97.47  & 98.51  \\
% Attacker\_4 & -     & -     & -     & 98.89  & 98.18  \\
% Attacker\_5 & -     & -     & -     & -      & 99.53  \\ 
\bottomrule
\end{tabular}%
}
\caption{Mirage with different attacker number.}
\label{Tab: different attacker number}
\vspace{-4 mm}
\end{table}

\subsection{Performance under varying non-IID degrees}

Tab.~\ref{Tab: different data distribution} exhibits the attack performance of Mirage under varying non-IID degrees. We set different sampling parameters $\alpha$ for Mirage, with larger values indicating a distribution that is closer to IID. The results indicate that Mirage can achieve stable performance across different non-IID degrees. As described in Sec.~\ref{Subsec: ID mapping enhancement}, the adversarial strategy may optimize the backdoor distribution to the margin of the target distribution. To address this issue, we introduce a constrained optimization method, tightening the backdoor distribution into the objective distribution and ensuring successful attacks across different distributions.

\subsection{Illustration of backdoor distribution}

\begin{figure*}
    \centering
    \setlength{\abovecaptionskip}{1 mm}    \includegraphics[width=0.98\textwidth]{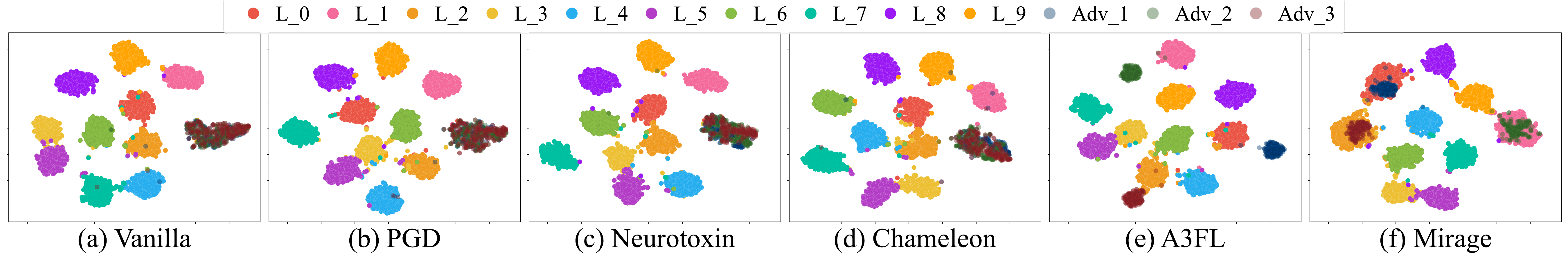}
    \vspace{-2 mm}
    \caption{Illustration of backdoor distribution. }
    
    \label{Fig: backdoor distribution}
    \vspace{-7 mm}
\end{figure*}

The preceding experiments leverage t-SNE to illustrate the backdoor distribution for Mirage and baseline methods. The results are in Fig.~\ref{Fig: backdoor distribution}. The backdoor distribution from Fig. 5(a) to Fig. 5(d) further corroborates our presentation in \ref{Sec: Introduction}, that different attackers may share similar neuron activation paths, resulting in their potential mitigation since their distribution is close to each other. On the other aspect, A3FL paves the activation path by adversarially unlearning the backdoor, which enhances their OOD mapping, producing clear boundaries between classes and tight clustering within each class. Although effective, this poses more threats to A3FL since the OOD-sample-based detection method can detect their backdoor models effectively.
Looking back to Mirage, which is pictured in Fig. 5(f), backdoor samples present in-distribution with the target clean distribution. The results further indicate that the proposed Mirage can construct the ID mapping between backdoor features and clean distributions of the target class. 

\subsection{Ablation study}

\begin{figure}
    \centering
    \setlength{\abovecaptionskip}{1 mm}    \includegraphics[width=0.75\linewidth]{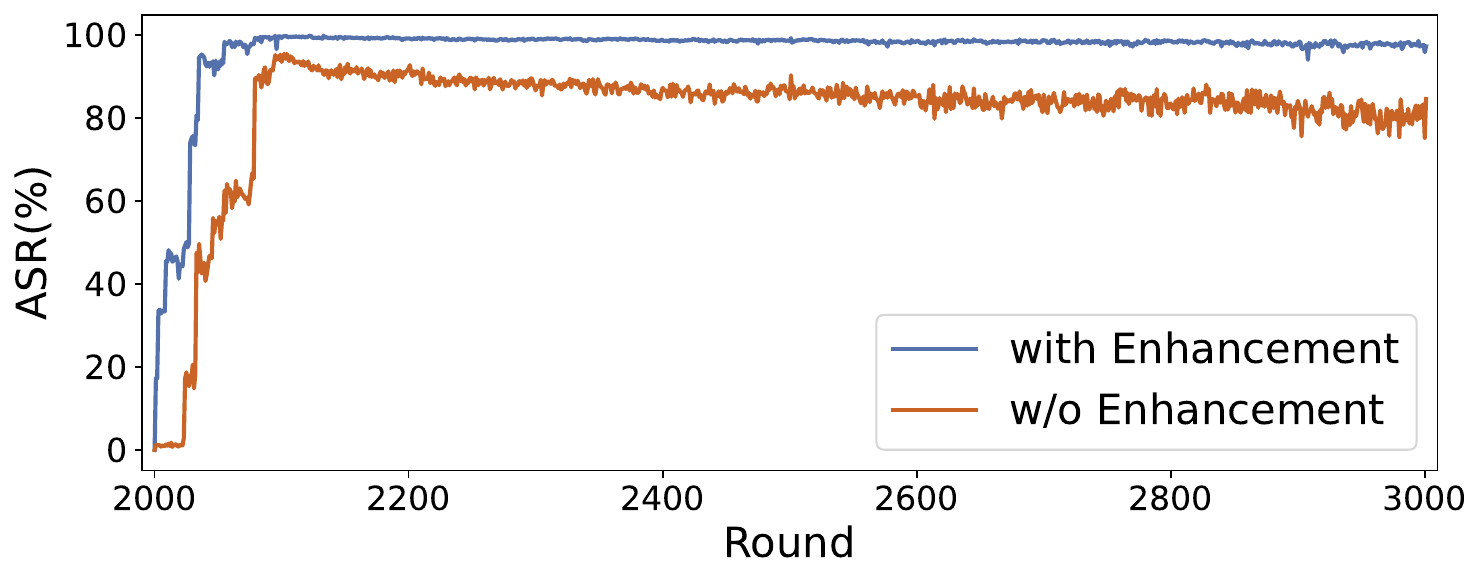}
    \vspace{-1.5 mm}
    \caption{Ablation study.}
    \vspace{-8 mm}
    \label{Fig: Ablation study}
\end{figure}

Sections~\ref{Subsec: ID mapping construction} and \ref{Subsec: ID mapping enhancement} propose two hybrid techniques to construct and enhance the ID mapping. Here, we conduct an ablation study to show the effects of different techniques for our method. The experiment adheres to the settings outlined in Section~\ref{Subsec: Lifespan}. Fig.~\ref{Fig: Ablation study} presents the average ASR for Mirage and Mirage without enhancement. We can see that introducing the ID mapping enhancement significantly improves Mirage's performance. As described in Section~\ref{Subsec: ID mapping enhancement}, although the adversarial strategy successfully constructs the ID mapping, the backdoors may not fully activate the clean path, making the backdoor distribution map to the target distribution's margin, rendering it ineffective and non-persistent. Therefore, we introduce the constrained optimization method to tighten the backdoor distribution, ensuring backdoor samples will not be misidentified as the target distribution changes.

\section{Discussion}
\subsection{Does the pixel trigger still effective?} 
\label{Subsec: pixel trigger}
% In the previous experiments, we adopted the blend trigger as the trigger pattern. Here, we conduct experiments with pixel trigger to evaluate its effectiveness. The results (presented in Appendix~\ref{Appendix: patched triggers}) shows that the pixel trigger remains effective. According to the t-SNE analysis of the feature attribution, the backdoor samples cannot aggregate together like the blend trigger. The main reason is that the pixel trigger can only activate a subset of neurons along the activation path, while other features typically activate the paths belonging to their original classes. However, it still constructs the ID mapping to the target distribution and uccessfully embeds the backdoors for different attackers.

In the previous experiments, we adopted the blend trigger as the trigger pattern. Here, we conduct experiments with pixel triggers to evaluate their effectiveness. The results (presented in Appendix~\ref{Appendix: patched triggers}) show that the pixel trigger remains effective. According to the t-SNE analysis of the feature attribution, the backdoor samples cannot aggregate together like the blend trigger. The main reason is that the pixel trigger can only activate a subset of neurons along the activation path, while other features typically activate the paths belonging to their original classes. However, it still constructs the ID mapping to the target distribution and successfully embeds the backdoors for different attackers.

% \begin{figure}
%     \centering
%     \includegraphics[width=0.5 \linewidth]{Figs/tsne/pixel_block_trigger_tsne.pdf}
%     \caption{The t-SNE of Mirage with pixel block as trigger pattern.}
%     \label{Fig: pixel block tsne}
%     \vspace{-5mm}
% \end{figure}

\subsection{One-shot Mirage} 
Mirage can perform a one-shot attack by increasing adversarial training epochs, allowing the trigger to directly simulate features of the target class, like adversarial samples. However, it does not construct persistent and effective ID backdoor mapping, resulting in low durability and side effects on the main tasks. Therefore, we enable the model to learn from the backdoor samples and construct effective and persistent ID mapping by injecting more rounds and fewer batches to achieve high-confidence classification.

\subsection{Potential defense methods}
\label{Subsec: Potential defense methods}
According to Section~\ref{Subsection: Attack performance}, we can see that although Indicator \cite{backdoorFL_bagdasaryan_2020}, the most powerful detection method for now, can detect at least one of the three A3FL attackers, the rest can still successfully inject the backdoor. This is because it constructs limited OOD mappings, succeeding only under the assumption that the attacker constructs similar OOD mappings. Therefore, it cannot be as effective against dynamic attacks (A3FL) as it is against static attacks (Chameleon, etc.). Furthermore, for A3FL, such misdetections are sufficient for the attacker to inject a persistent backdoor into the global model. Thus, following the works with foresight \cite{NBA_nguyen_2024, SEN_li_2024}, we urge researchers to focus on MBA and expand their efforts to defend against it.

% detection-based defense methods are generally weak in MBA. Although BackdoorIndicator \cite{backdoorFL_bagdasaryan_2020}, the most powerful detection method for now, can detect at least one of three A3FL attackers, the rest can still successfully inject the backdoor. We consider the reason is that previous works may focus on the SBA, result in some attackers may bypass the detection under the cover of “unfortunate” attackers. One thing should be consider is, when A3FL meets BackdoorIndicator, 

% We consider the reason behind this to be that current defense methods work under the assumption that there is only one attacker. Specifically, BackdoorIndicator, the most powerful detection method for now, only considers the label with the largest ASR on their OOD samples, resulting in it only rejecting one attacker update but setting others free for attacking. Thus, following the works with foresight \cite{NBA_nguyen_2024, SEN_li_2024}, we call on researchers to pay attention to MBA and extend their works to defend against it. 

Our proposed Mirage does not follow the traditional backdoor activation scheme, making it challenging for previous work to resist. This motivates us to seek solutions to address this threat to real-world FL applications. The advanced explainable method may effectively detect our attack, however, detecting all backdoored models remains challenging. Input detection may also identify our attack, as clean samples differ from backdoored samples.

% \subsection{Attacker collusion}

\section{Conclusion}
\label{Sec: Summary and Future Work}
In this paper, we uncover the inherent constraints of SBAs when extending them to MBA scenarios: Non-cooperative attackers will exclude others as they construct similar backdoor mappings. Based on this observation, we introduce Mirage, the first non-cooperative MBA strategy in FL. Mirage ensures the backdoor functions of different attackers by constructing effective and persistent ID backdoor mappings. Specifically, an adversarial trigger optimization method is introduced to construct the ID mapping between backdoor features and target distributions. Subsequently, Mirage leverages a constrained optimization method to tighten the ID mapping for persistence. Our comprehensive experiments demonstrate that Mirage outperforms existing backdoor attacks across different settings. For future work, investigating defense methods will be our primary focus.
% \section{Acknowledgement}

\noindent \textbf{Acknowledgement}

This work was supported in part by the National Natural Science Foundation of China under Grant(No. 62172215) and in part by the A3 Foresight Program of NSFC (Grant No. 62061146002)

\clearpage
{
    \small
    \bibliographystyle{ieeenat_fullname}
    \bibliography{main}
}
\renewcommand\thesection{\Alph{section}}
\setcounter{section}{0}
\clearpage
\setcounter{page}{1}
\maketitlesupplementary

\section{Extended Experiment and Analysis}
We provide an extended experiment to further demonstrate the generality of the conflict in MBA scenarios. In addition to this, we give a simple theoretical analysis of confliction.

\subsection{Experiments with different attack methods}
\label{Appendix: extend experiments}
\begin{figure}[htb!]
    \centering
    \includegraphics[width=\linewidth]{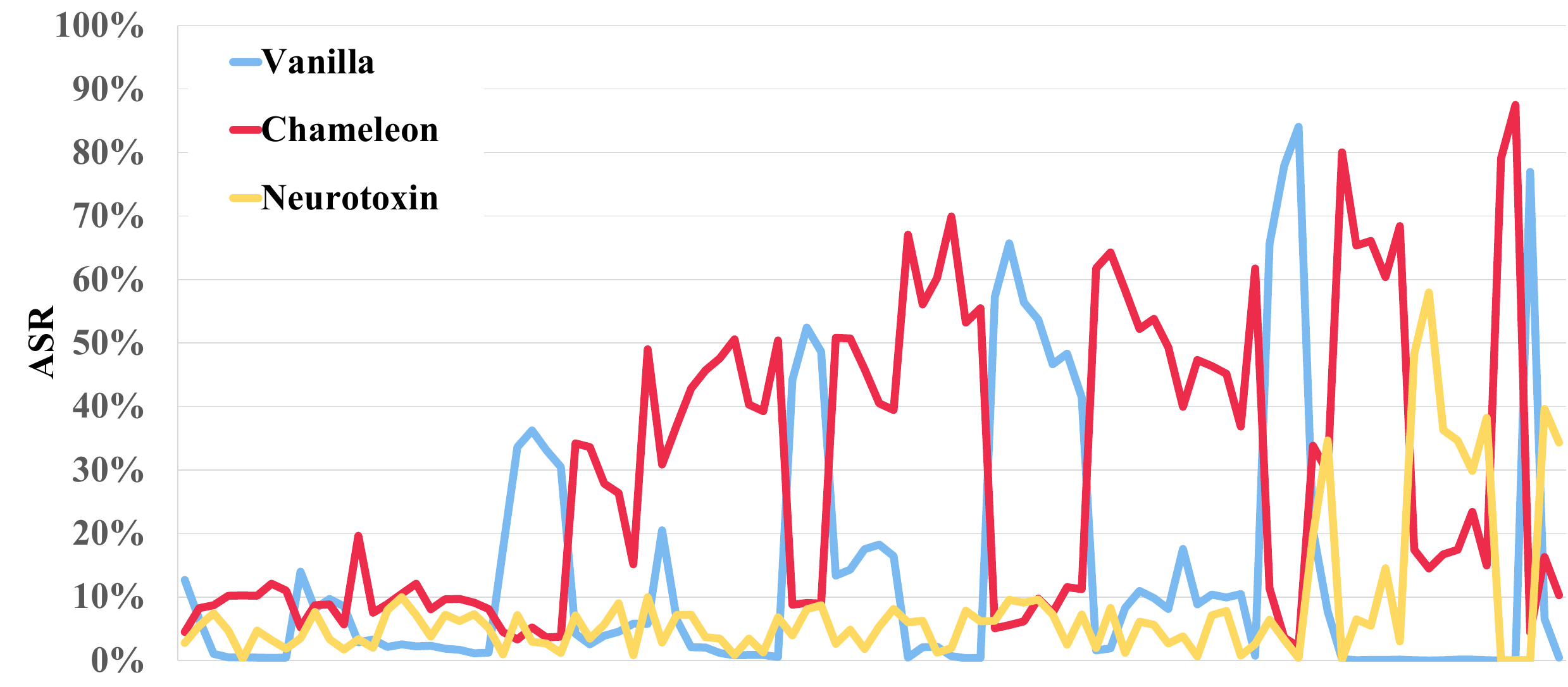}
    \caption{ASRs of different attack algorithms.}
    \label{fig: different_algo_conflict}
    \vspace{-8mm}
\end{figure}

In Fig.~\ref{Fig: toy_experiment}, we illustrate the conflict arising when two attackers employ the same attack method (Vanilla). Here, we further conduct an extend experiment using three distinct attack methods under non-collusive conditions. The ASR curves are depicted in Fig.~\ref{fig: different_algo_conflict}. These experimental results validate our motivation, indicating that different attackers will indeed conflict in MBA scenarios. Furthermore, NBA \cite{NBA_nguyen_2024} has also observed this issue, yet they did not provide an explanation or an effective solution. In this paper, we conduct an in-depth investigation to uncover the inherent constraints of exclusion and propose an effective backdoor attack method to address these constraints.

% Here, we further conduct an toy experiments with three different attack methods under non-collusive conditions, the ASR curves are shown in \ref{fig: different_algo_conflict}. Such experimental results further show that our motivation holds, i.e., different attackers will conflict with each other in MBA scenarios. Besides, NBA \cite{NBA_nguyen_2024} also observed this problem, but they failed to explain the reason and propose an effective solution. In this paper, we conduct an in-depth investigation to uncover the inherent constraints of the exclusion and further propose an effective backdoor attackers address such constraints. 

% Figure~\ref{Fig: toy_experiment} illustrates the conflict among two attackers using the same attack method (Vanilla). We further conduct a toy experiment with three different attack methods. The corresponding ASR curves are presented in Figure~\ref{fig: different_algo_conflict}. These results reinforce our claim that different attackers will indeed experience conflicts in MBA scenarios. Beside,  

\subsection{Theoretical Analysis}
The inference of a CNN model can be represented as $\theta_c(\theta_f(x))$, where $\theta_c(\cdot)$ is a classifier and $\theta_f(\cdot)$ is the feature extractor. As we present in Fig.~\ref{Fig: toy_experiment} t-SNE, if two attackers (R and B) construct similar Out-of-Distribution (OOD) mapping for their backdoor function, their backdoor samples ($x_R$ and $x_B$) exhibit overlapping distributions in the feature space, i.e., $\theta_f(x_R)$ and $\theta_f(x_B)$ are similar. Such similarity renders the backdoor samples of them indistinguishable. So, the predictions depend on which attacker dominates the model, a factor that is inherently uncontrollable (see Fig.~\ref{Fig: toy_experiment}, where the changing ASRs indicate ongoing conflicts). To prevent R and B from constructing similar backdoor mappings, we construct In-Distribution (ID) mapping. Therefore, $\theta_f(x_R)$ and $\theta_f(x_B)$ will be mapped to the clean distribution of their target classes (Fig.~\ref{Fig: 2_c}) to avoid conflict under non-collusive conditions.

\section{Algorithm Outline}
\label{Appendix: Algorithm Outline}
We describe the process of Mirage as follows: 

At $t$-th FL round, the attacker (assuming the index of this attacker is $i$) is selected by the server and receives the latest FL global model $\theta_t$. Lines 4-5 train the detector based on the feature extractor of the current global model and the trigger pattern to discriminate whether a sample is OOD sample. Lines 7-10 optimize the trigger pattern based on the detector loss (proposed in Section~\ref{Subsec: ID mapping construction}) and the enhancement loss (provided in Section~\ref{Subsec: ID mapping enhancement}) for one batch of data. Lines 14-17 train the local model on the poisoned dataset and upload the local updates to the server.

\section{Datasets Details}
\label{Appendix: datasets}
In the experimental evaluations, we leverage three computer vision datasets: CIFAR-10, CIFAR-100 \cite{Cifar_Krizhevsky_2009}, and GTSRB to evaluate the performance of our proposed method.

 \textbf{CIFAR-10}: The CIFAR-10 dataset consists of 60,000 32x32 color images in 10 classes, including dogs, cats, and cars. For each class, there are a total of 6,000 samples, with 5,000 for training and 1,000 for testing.

\textbf{CIFAR-100}: The CIFAR-100 dataset is similar to CIFAR-10, except it has 100 classes containing 600 images each, with 500 training images and 100 testing images. Additionally, these 100 classes can be grouped into 20 superclasses, such as aquarium fish, flatfish, ray, shark, and trout, which can be grouped into the superclass "fish." In this paper, we use the 100 classes rather than the 20 superclasses for evaluations.

\textbf{GTSRB}: The German Traffic Sign Recognition Benchmark (GTSRB) contains 43 classes of traffic signs, divided into 39,209 training images and 12,630 test images, each with a size of 32x32 pixels.

\section{Different Attacker Numbers}
\label{Appendix: Different Attacker Numbers}

In the previous evaluations, we demonstrated the attack performance of Mirage under varying numbers of attackers, ranging from 1 to 5, on CIFAR-10. Additionally, we conducted experiments on two other datasets, and the results are presented in Table~\ref{Tab: different attacker on cifar100 and GTSRB}. The experimental results across different datasets are consistent, indicating that Mirage exhibits high usability with varying numbers of attackers and does not induce potential infighting among them.

% \begin{tabular}{ccccccc}
% \toprule
% \textbf{Dataset}          & Metric & \textbf{N=1} & \textbf{N=2} & \textbf{N=3} & \textbf{N=4} & \textbf{N=5} \\ \midrule
% \multirow{3}{*}{CIFAR100} & Acc                      & 71.7       & 72.04      & 71.65      & 72.05      & 71.64      \\
%                           & ASR                      & 99.82      & 99.5       & 99.05      & 99.26      & 99.21      \\
%                           & GAP                      & 0.00          & 0.32       & 0.77       & 1.43       & 1.14       \\ \midrule
% \multirow{3}{*}{GTSRB}    & Acc                      & 96.55      & 96.68      & 96.97      & 96.79      & 96.64      \\
%                           & ASR                      & 99.73      & 99.61      & 99.73      & 99.19      & 99.58      \\
%                           & GAP                      & 0.00          & 0.43       & 0.48       & 2.69       & 1.48 \\     
% \bottomrule
% \end{tabular}%
% }
% \caption{Performances for different attack number N in CIFAR100 and GTSRB.}
% \vspace{-3mm}
% \label{Tab: different attacker on cifar100 and GTSRB}
% \end{table}

\section{Patched Triggers}
\label{Appendix: patched triggers}
\begin{figure}[htb!]
    \centering
    \includegraphics[width=0.5 \linewidth]{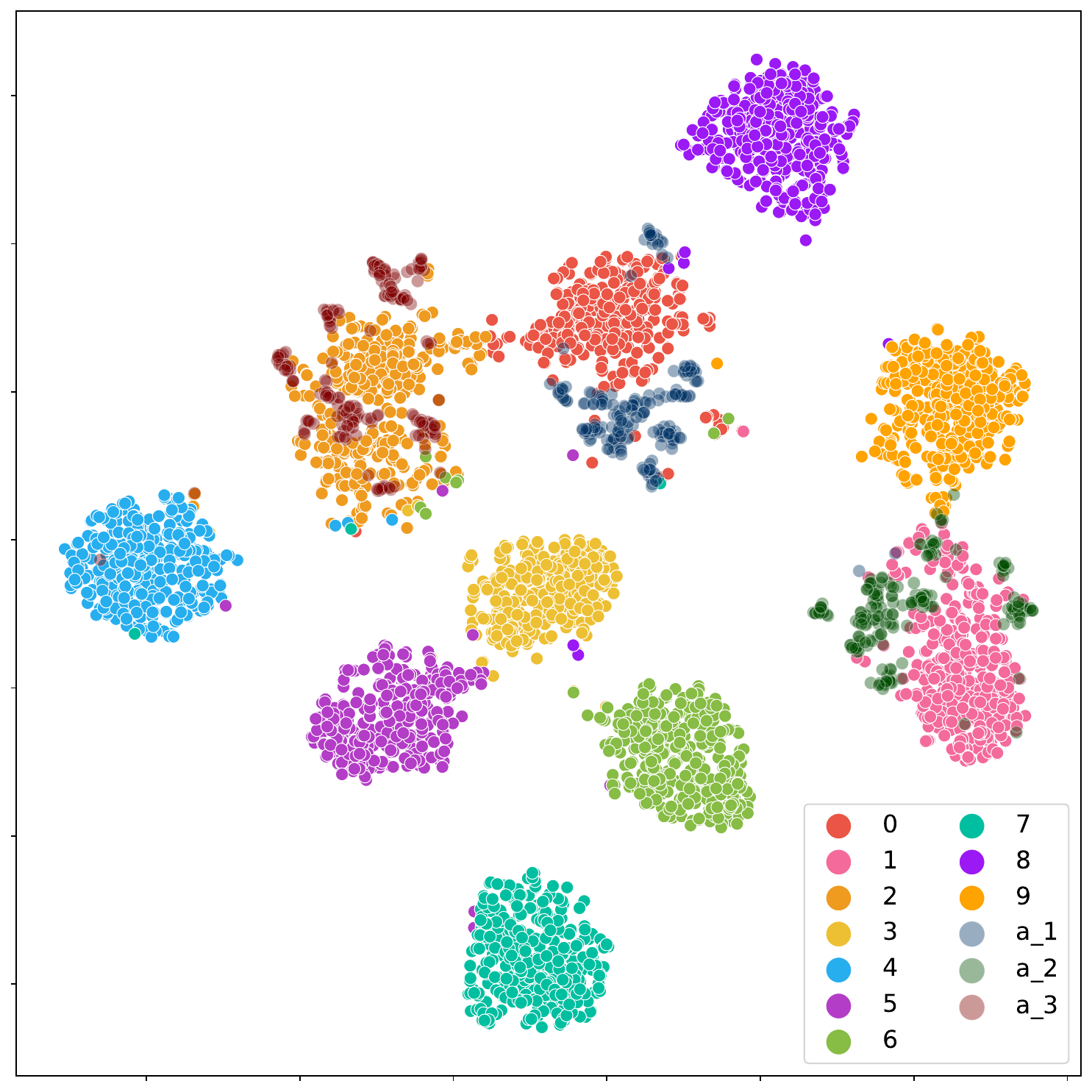}
    \caption{The t-SNE of Mirage with pixel block as trigger pattern.}
    \label{Fig: pixel block tsne}
    \vspace{-4mm}
\end{figure}

In Section~\ref{Sec: Experimental Evaluation}, we evaluate the effectiveness of Mirage based on blend triggers \cite{Blend_chen_2017}. Consequently, we also use a square patch as the trigger for Mirage. In the implementation, we set the block size to 5x5 and applied it to the top-left corner of each sample. Aside from that, we do not change any other parameters in the default settings. The t-SNE results are presented in Fig~\ref{Fig: pixel block tsne}, and the discussion is provided in Section~\ref{Subsec: pixel trigger}. The performance of the patched triggers are presented in Table~\ref{Tab: patched triggers} that the trigger pattern has a slight influence on the ASRs, yet acceptable.

\begin{table}[]
\centering
\resizebox{0.6\linewidth}{!}{%
\begin{tabular}{c|ccc}
\toprule
    & CIFAR10 & CIFAR100 & GTSRB   \\ \midrule
Acc & 92.40\% & 71.28\%  & 96.66\% \\
ASR & 95.67\% & 96.06\%  & 94.63\% \\ \bottomrule
\end{tabular}
}
\caption{Performances of patched triggers.}
\label{Tab: patched triggers}
\end{table}

\begin{table}[]
\centering
\resizebox{\linewidth}{!}{%
\begin{tabular}{c|cccccc}
\toprule
\textbf{Dataset}          & \textbf{Attacker Number} & \textbf{$N=1$} & \textbf{$N=2$} & \textbf{$N=3$} & \textbf{$N=4$} & \textbf{$N=5$} \\ \midrule
\multirow{7}{*}{CIFAR10}
& Acc ($\uparrow$)        & 92.37 & 92.53 & 92.16 & 92.11  & 92.12  \\
&ASR ($\uparrow$)        & 99.06 & 99.15 & 98.80 & 98.375 & 98.484 \\  \cmidrule{2-7}
&Attacker\_1 & 99.06 & 99.39 & 99.01 & 98.71  & 98.66  \\
&Attacker\_2 & -     & 98.9  & 98.63 & 98.43  & 97.54  \\
&Attacker\_3 & -     & -     & 98.77 & 97.47  & 98.51  \\
&Attacker\_4 & -     & -     & -     & 98.89  & 98.18  \\
&Attacker\_5 & -     & -     & -     & -      & 99.53  \\ \midrule
\multirow{7}{*}{CIFAR100} & Acc ($\uparrow$)                      & 71.70      & 72.04      & 71.65      & 72.05      & 71.64      \\
                          & ASR ($\uparrow$)                      & 99.82      & 99.50      & 99.05      & 99.26      & 99.21      \\ \cmidrule{2-7} 
                          & Attacker\_1              & 99.82      & 99.66      & 99.47      & 99.91      & 99.79      \\
                          & Attacker\_2              & -          & 99.34      & 98.98      & 98.95      & 99.21      \\
                          & Attacker\_3              & -          & -          & 98.70      & 99.71      & 99.24      \\
                          & Attacker\_4              & -          & -          & -          & 98.48      & 98.65      \\
                          & Attacker\_5              & -          & -          & -          & -          & 99.16      \\ \midrule
\multirow{8}{*}{GTSRB}    & Acc ($\uparrow$)                     & 96.55      & 96.68      & 96.97      & 96.79      & 96.64      \\
                          & ASR ($\uparrow$)                     & 99.73      & 99.61      & 99.73      & 99.19      & 99.58

                          \\ \cmidrule{2-7} 
                          & Attacker\_1              & 99.73      & 99.82      & 99.79      & 99.86      & 99.98      \\
                          
                          & Attacker\_2              & -          & 99.39      & 99.46      & 99.60      & 99.73      \\
                          & Attacker\_3              & -          & -          & 99.94      & 100.00     & 99.98      \\
                          & Attacker\_4              & -          & -          & -          & 97.31      & 98.50      \\
                          & Attacker\_5              & -          & -          & -          & -          & 99.73      \\ \bottomrule
\end{tabular}
}%
\caption{Performance for different attack numbers $N$ across three datasets. The Acc and ASR represent the averages for $N$ attackers, and the detailed ASR is provided in the following items.}
\vspace{-3mm}
\label{Tab: different attacker on cifar100 and GTSRB}
\end{table}

\section{Trigger Visualization}

\begin{figure}
\centering
% 第一行图像
\begin{minipage}[b]{0.28\linewidth}
    \centering
    \includegraphics[width=\linewidth]{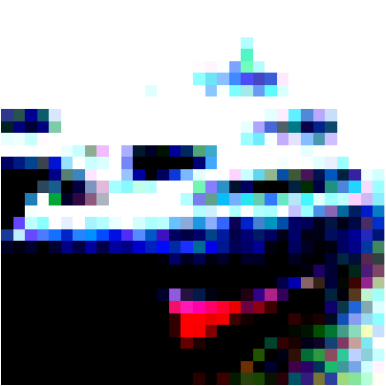}
    \subcaption{Vanilla\_0\_sample}
    \label{trigger_vanilla_0}
\end{minipage}%   % 注意这里的 % 符号
\hspace{0.02\textwidth}%
\begin{minipage}[b]{0.28\linewidth}
    \centering
    \includegraphics[width=\linewidth]{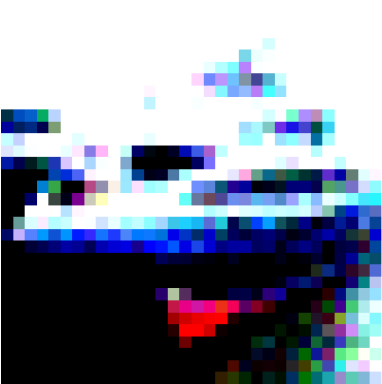}
    \subcaption{Vanilla\_1\_sample}
\end{minipage}%
\hspace{0.02\textwidth}%
\begin{minipage}[b]{0.28\linewidth}
    \centering
    \includegraphics[width=\linewidth]{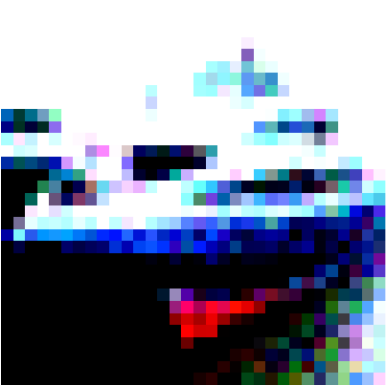}
    \subcaption{Vanilla\_2\_sample}

\end{minipage}

\vspace{0.5em}% 控制两行之间的垂直间距

% 第二行图像
\begin{minipage}[b]{0.28\linewidth}
    \centering
    \includegraphics[width=\linewidth]{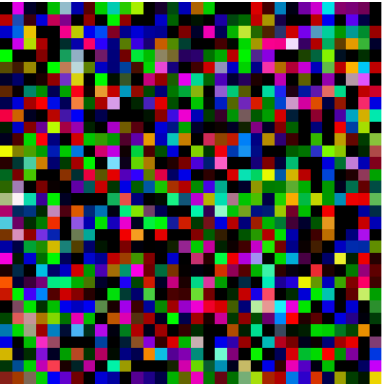}
    \subcaption{Vanilla\_0\_trigger}
\end{minipage}%
\hspace{0.02\textwidth}%
\begin{minipage}[b]{0.28\linewidth}
    \centering
    \includegraphics[width=\linewidth]{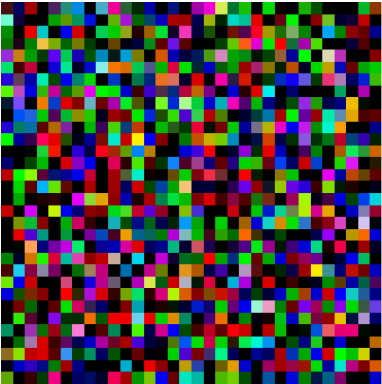}
    \subcaption{Vanilla\_1\_trigger}
\end{minipage}%
\hspace{0.02\textwidth}%
\begin{minipage}[b]{0.28\linewidth}
    \centering
    \includegraphics[width=\linewidth]{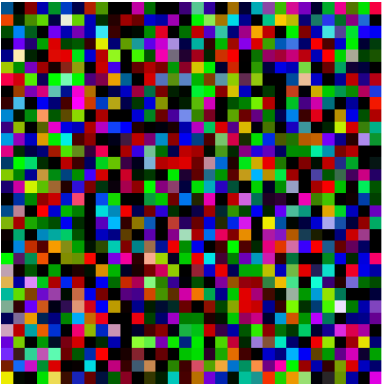}
    \subcaption{Vanilla\_2\_trigger}
\end{minipage}
\vspace{-2mm}
\caption{Vanilla Visualization}
\label{fig: triggers_Vanilla}
\vspace{-4mm}
\end{figure}

\begin{figure}
\centering
% 第一行图像
\begin{minipage}[b]{0.28\linewidth}
    \centering
    \includegraphics[width=\linewidth]{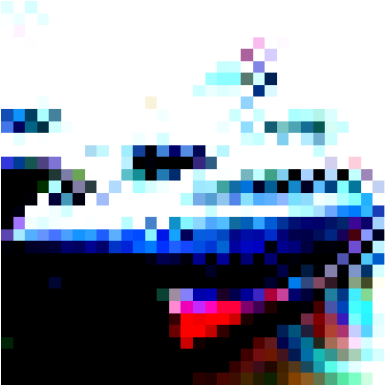}
    \subcaption{A3FL\_0\_sample}
\end{minipage}%   % 注意这里的 % 符号
\hspace{0.02\textwidth}%
\begin{minipage}[b]{0.28\linewidth}
    \centering
    \includegraphics[width=\linewidth]{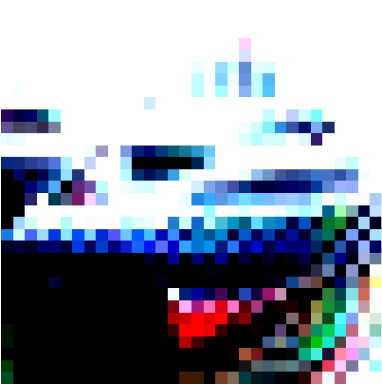}
    \subcaption{A3FL\_1\_sample}
\end{minipage}%
\hspace{0.02\textwidth}%
\begin{minipage}[b]{0.28\linewidth}
    \centering
    \includegraphics[width=\linewidth]{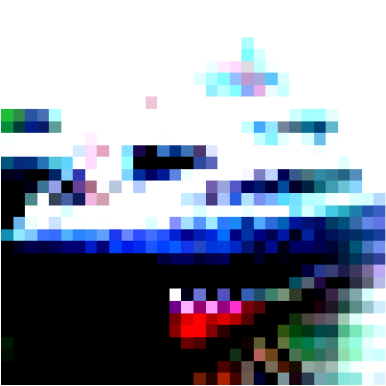}
    \subcaption{A3FL\_2\_sample}
\end{minipage}

\vspace{0.5em}% 控制两行之间的垂直间距

% 第二行图像
\begin{minipage}[b]{0.28\linewidth}
    \centering
    \includegraphics[width=\linewidth]{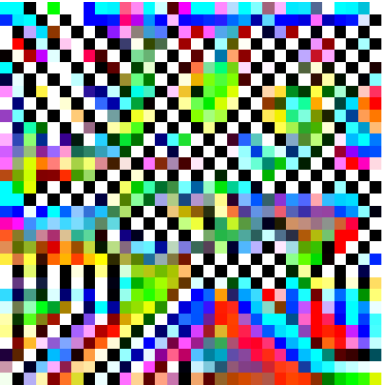}
    \subcaption{A3FL\_0\_trigger}
\end{minipage}%
\hspace{0.02\textwidth}%
\begin{minipage}[b]{0.28\linewidth}
    \centering
    \includegraphics[width=\linewidth]{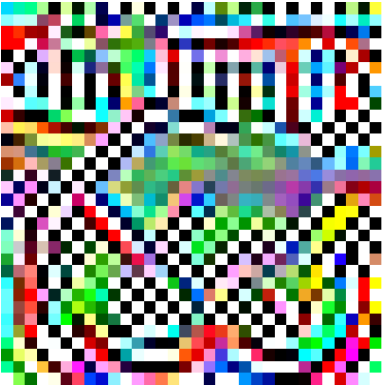}
    \subcaption{A3FL\_1\_trigger}
\end{minipage}%
\hspace{0.02\textwidth}%
\begin{minipage}[b]{0.28\linewidth}
    \centering
    \includegraphics[width=\linewidth]{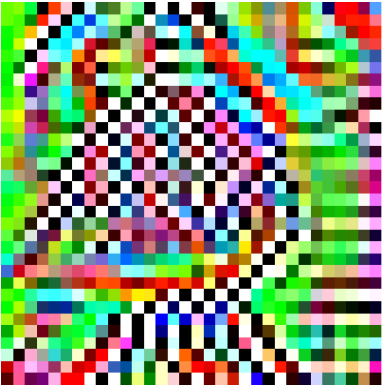}
    \subcaption{A3FL\_2\_trigger}
\end{minipage}
\vspace{-2mm}
\caption{A3FL Visualization}
\label{fig: triggers_A3FL}
\vspace{-4mm}
\end{figure}

\begin{figure}
\centering
% 第一行图像
\begin{minipage}[b]{0.28\linewidth}
    \centering
    \includegraphics[width=\linewidth]{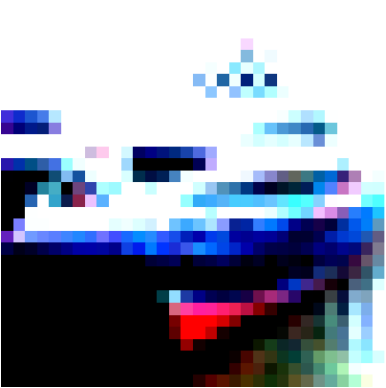}
    \subcaption{Mirage\_0\_sample}
\end{minipage}%   % 注意这里的 % 符号
\hspace{0.02\textwidth}%
\begin{minipage}[b]{0.28\linewidth}
    \centering
    \includegraphics[width=\linewidth]{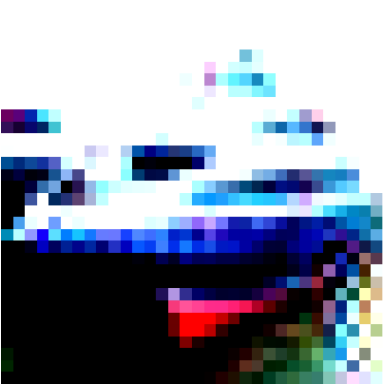}
    \subcaption{Mirage\_1\_sample}
\end{minipage}%
\hspace{0.02\textwidth}%
\begin{minipage}[b]{0.28\linewidth}
    \centering
    \includegraphics[width=\linewidth]{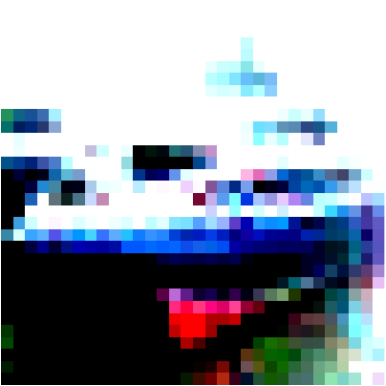}
    \subcaption{Mirage\_2\_sample}
\end{minipage}

\vspace{0.5em}% 控制两行之间的垂直间距

% 第二行图像
\begin{minipage}[b]{0.28\linewidth}
    \centering
    \includegraphics[width=\linewidth]{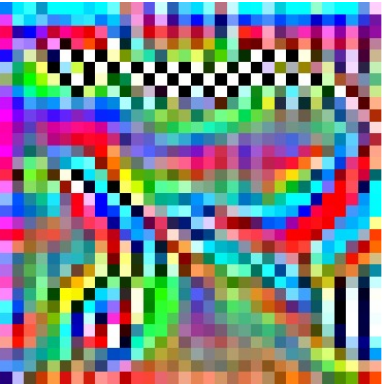}
    \subcaption{Mirage\_0\_trigger}
\end{minipage}%
\hspace{0.02\textwidth}%
\begin{minipage}[b]{0.28\linewidth}
    \centering
    \includegraphics[width=\linewidth]{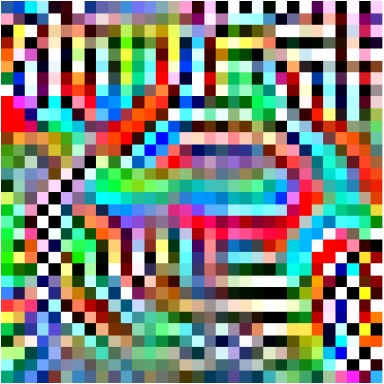}
    \subcaption{Mirage\_1\_trigger}
\end{minipage}%
\hspace{0.02\textwidth}%
\begin{minipage}[b]{0.28\linewidth}
    \centering
    \includegraphics[width=\linewidth]{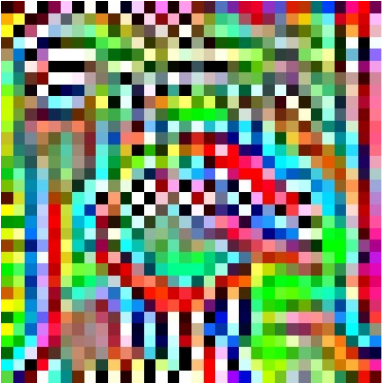}
    \subcaption{Mirage\_2\_trigger}
\end{minipage}
\vspace{-2mm}
\caption{Mirage Visualization}
\label{fig: triggers_Mirage}
\vspace{-4mm}
\end{figure}

We present the triggers and examples of backdoor samples of Mirage, A3FL and Vanilla in Fig.~\ref{fig: triggers_Vanilla}, Fig.~\ref{fig: triggers_A3FL} and Fig.~\ref{fig: triggers_Mirage}. The visualization results for PGD, Neurotoxin, and Chameleon are omitted since these attacks use static triggers, differing solely in their training methodologies. Consequently, their triggers align with Vanilla's setup, meaning the same attacker uses identical triggers across all static-trigger attack methods. For instance, Attacker 0 uses the trigger shown in Fig.~\ref{trigger_vanilla_0} consistently across Vanilla, PGD, Neurotoxin, and Chameleon, as do Attackers 1 and 2.

\end{document}